\documentclass[12pt,a4paper]{article}
\usepackage{a4wide}
\usepackage[centertags]{amsmath}
\usepackage{amssymb}
\usepackage{cite}
\usepackage{ifpdf}
\usepackage{color}
\usepackage{graphicx}
\newcommand{\ra}{\rightarrow}

\newcommand{\cdt}{\!\cdot\!}
\newcommand{\vp}{\vspace{0.5cm}}
\newcommand{\prn}[1]{\left ( #1 \right )}
\newcommand{\brc}[1]{\left\{ #1 \right\}}
\newcommand{\brk}[1]{\left [ #1 \right ]}
\newcommand{\abs}[1]{\left | #1 \right |}
\newcommand{\diff}[3][\rule{0mm}{0mm}]{\frac{\mathrm{d}^{#1} #2}{\mathrm{d}{#3}^{#1}}}
\newcommand{\pdiff}[3][\rule{0mm}{0mm}]{\frac{\partial^{#1} #2}{\partial {#3}^{#1}}}
\newcommand{\pdiffc}[3][\rule{0mm}{0mm}]{\left (\frac{\partial #2}{\partial {#3}}\right )_{\!\!#1}}
\newcommand{\dr}{\mathrm{d}}
\newcommand{\e}{\mathrm{e}}
\newcommand{\preprintno}[3]{\hfill\raisebox{#1}[0cm][0cm]{
\begin{minipage}[t]{#2}\begin{flushright} #3 \end{flushright}\end{minipage}}
\vspace*{-\baselinestretch\baselineskip}}
\newcommand{\CO}{\mathcal{O}}
\newcommand{\CN}{\mathcal{N}}

\newcommand{\half}{\frac{1}{2}}
\newcommand{\p}{\partial}
\newcommand{\apm}{{\alpha^{\prime}}}
\newcommand{\rz}{\rho_0}
\newcommand{\tA}{\hat{\mathrm{A}}}
\newcommand{\tB}{\hat{\mathrm{B}}}
\newcommand{\tC}{\hat{\mathrm{C}}}
\newcommand{\tOr}{\hat{\mathrm{O}}}
\newcommand{\tE}{\widetilde{E}}
\newcommand{\tL}{\widetilde{L}}
\newcommand{\tS}{\widetilde{S}}
\newcommand{\tT}{\widetilde{T}}
\newcommand{\tO}{\widetilde{\Omega}}
\newcommand{\eos}{\alpha}
\newcommand{\tc}{\mathcal{T_\mathrm{c}}}
\newcommand{\tloc}{\mathcal{T}}
\newcommand{\floc}{\mathcal{F}}
\newcommand{\eloc}{\mathcal{E}}
\newcommand{\sloc}{\mathcal{S}}

\newcommand{\ri}{r_\mathrm{i}}
\newcommand{\ro}{r_\mathrm{o}}
\newcommand{\vi}{v_\mathrm{i}}
\newcommand{\vo}{v_\mathrm{o}}
\newcommand{\z}{\tilde{r}}
\newcommand{\zi}{\z_\mathrm{i}}
\newcommand{\zo}{\z_\mathrm{o}}
\newcommand{\tw}{\widetilde{\omega}}
\newcommand{\g}{\tilde h}
\newcommand{\vot}{\tilde{v}_\mathrm{o}}
\newcommand{\vit}{\tilde{v}_\mathrm{i}}
\newcommand{\vm}{v_\mathrm{m}}

\title{Plasmarings as dual black rings}
\author{Subhaneil Lahiri$^{(a,b)}$ and
        Shiraz Minwalla$^{(a)}$
\\
\small{\emph{$^{(a)}$Department of Theoretical Physics,
                   Tata Institute of Fundamental Research,}}\\
\small{\emph{Homi Bhabha Rd, Mumbai 400005, India}}\\
\small{\emph{$^{(b)}$Jefferson Physical Laboratory,
                   Harvard University, Cambridge MA 02138, USA}}
}

\begin{document}

\maketitle

\preprintno{8cm}{6cm}{
    \texttt{arXiv:0705.3404 [hep-th]}
}


\begin{abstract}
We construct solutions to the relativistic Navier-Stokes equations
that describe the long wavelength collective dynamics of the
deconfined plasma phase of $\CN=4$ Yang Mills theory compactified
down to $d=3$ on a Scherk-Schwarz circle and higher dimensional generalisations. Our solutions are stationary, axially symmetric spinning balls and rings of plasma. These solutions, which are dual to (yet to be constructed) rotating black holes and black rings in Scherk-Schwarz compactified AdS$_5$ and AdS$_6$, and have properties that are qualitatively similar to those of black holes and black rings in flat five dimensional supergravity.
\end{abstract}

\tableofcontents



\section{Introduction}\label{sec:intro}

A particularly interesting entry in the dictionary between gauge
theory and gravity links deconfined or `gluon plasma' phase of Yang
Mills theory to black branes and black holes in gravity. In this
paper we study aspects of this connection in the context of specific
examples. In most of this paper we study  $d=4$, $SU(N)$, $\CN=4$
Yang-Mills at 't~Hooft coupling $g^2_{YM}N=\lambda$, compactified on
a Scherk-Schwarz $S^1$ (the remaining $2+1$ dimensions are non
compact). The low energy dynamics of this theory is that of a $2+1$
dimensional Yang-Mills system that undergoes deconfining phase
transition at a finite temperature \cite{Witten:1998zw}. At large
$N$ and strong 't~Hooft coupling this system admits supergravity
dual description; the low temperature confining phase is dual to a
gas of IIB supergravitons on the so called AdS soliton background
\cite{Witten:1998zw}
\begin{equation}\label{deconfmet:eq}
  \dr s^2 = L^2 \apm \left( \e^{2u}
       \left( -\dr t^2 + T_{2 \pi}(u)\, \dr\theta^2 + \dr w_i^2 \right)
       + \frac{1}{T_{2 \pi}(u)}\, \dr u^2 \right),
\end{equation}
where $i= 1,\cdots, 2$, $\theta\sim\theta+2\pi$, $L^2=
\sqrt{\lambda}$ and\footnote{Notice that, at large $u$,
$T_x(u)\simeq 1$, so \eqref{deconfmet:eq} reduces to AdS$_{d+2}$ in
Poincar\'e-patch coordinates, with $u$ as the radial scale
coordinate, and with one of the spatial boundary coordinates,
$\theta$, compactified on a circle (the remaining boundary
coordinates, $w_i$ and $t$, remain non-compact).}
\begin{equation}\label{AdSSchII:eq}
 T_{x}(u) = 1 - \left( \frac{x}{\pi} \,\e^u \right)^{-4}.
\end{equation}

The high temperature phase of the same system (at temperature
$\tloc=1/\beta$) is dual to the the black brane
\begin{equation} \label{AdSSchIb:eq}
  \dr s^2 = L^2 \apm \left( \e^{2 u}
     \left( -T_{\beta}(u)\, \dr t^2 + \dr\theta^2 + \dr w_i^2 \right)
     + \frac{1}{T_{\beta}(u)}\, \dr u^2 \right) .
\end{equation}
The thermodynamics of the high temperature phase are determined in
the bulk description by the usual constitutive equations of black
brane thermodynamics \cite{Aharony:2005bm}
\begin{equation}\label{equationofstate:eq}
P=-f=\frac{\pi^2 N^2} {8 \tc}\left( \tloc^4-\tc^4 \right).
\end{equation}
For $\tloc>\tc$ this free energy is negative, and so (in the large
$N$ limit) is smaller than the $\CO(1)$ free energy of the
`confined' gas of gravitons. Consequently, the system undergoes a
deconfinement phase transition at temperature $\tc$.\footnote{$\tc =
1/2 \pi$ in the dimensionless units of \eqref{AdSSchIb:eq}}

Just as the mean equilibrium properties of the deconfined phase are
well described by the equations of thermodynamics, the statistically
averaged near-equilibrium dynamics of this phase is governed by the
equations of fluid dynamics - the relativistic generalisation of the
Navier-Stokes equations. These equations accurately describe the
time evolution of fluid configurations whose space time derivatives
are all small in units of the mean free path, which is of the same
order as the mass gap of the theory \cite{Son:2007vk,
Aharony:2005bm}. The same equations, augmented by appropriate
surface terms, may also be used to study the dynamics of large lumps
of plasma localised in the gauge theory vacuum.

The properties of the surface that separates the plasma from the
vacuum, may be studied in the context of the simplest plasma profile
with a surface; a configuration in which half of space, $x<0$, is
filled with the plasma. The surface at $x=0$ is a domain wall that
separates the plasma from the vacuum. The net force on this domain
wall vanishes (and so the system is in equilibrium)  when the plasma
that fills $x<0$ has vanishing pressure, i.e. at $\tloc=\tc$ in the
large $N$ limit. The bulk gravity dual of this solution was
constructed numerically in \cite{Aharony:2005bm}; this configuration
interpolates between the black brane at $\tloc=\tc$ for $x<0$ and
the vacuum at $x>0$, via a domain wall. The thickness and surface
tension of this domain wall may be read off from this gravitational
solutions, and  were estimated, in \cite{Aharony:2005bm} at
approximately $6\times \frac{1}{2\pi \tc}$ and $\sigma=2.0 \times
\frac{\pi^2 N^2 \tc^2}{2}$ .

More generally, one would expect a finite lump of plasma that
evolves according to the relativistic Navier-Stokes equations map in
the bulk to a `black hole' that evolves according to the Einstein
equations. Provided all length scales in the plasma solution are
small compared to the gauge theory mass gap (which is of the same
order as the domain wall thickness), the dual bulk solution is well
approximated by a superposition of patches of the black brane
solution (with temperature varying across the patches) in the bulk
and patches of the domain wall solution described in the previous
paragraph. It follows (at least for stationary solutions) that the 3
dimensional black hole horizon topology (at any given time) is given
by an $S^1$ (physically this is the $\theta$ circle) fibred over the
two dimensional fluid configuration at the same time, subject to the
condition that the $S^1$ contracts at all fluid boundaries.
Consequently, fluid configurations with different topologies yield
bulk dual black hole configurations with distinct horizon
topologies. We will return to this point below.

This paper is devoted to a detailed study of certain `stationary'
configurations of the plasma fluid; i.e. time independent, steady
state solutions to the relativistic Navier-Stokes equations. The
simplest configurations of this sort was studied already in
\cite{Aharony:2005bm}; the plasmaball is a static, spherically
symmetric lump of fluid at constant local pressure $P$ with $P=
\sigma /R$ where $R$ is the radius of the lump and $\sigma$ its
surface tension. In this paper we study the more intricate spinning
lumps of stationary fluid. These lumps carry angular momentum in
addition to their mass.

It turns out that the relativistic Navier-Stokes equations admit two
distinct classes of solutions of these sort. The first class of
solution is a simple deformation of the static plasmaball; it is
given by plasmaballs that spin at a constant angular velocity. The
centripetal force needed to keep the configuration rotating in this
solution is provided by a pressure gradient. The local plasma
pressure (and hence local temperature and density) decreases from
the edge (where it is a positive number set by the radius, surface
tension and rotation speed) to the centre. As large enough angular
velocity the pressure goes sufficiently negative in the core of the
solution to allow for a second kind of solution of these equations;
an annulus of plasma fluid rotating at constant angular velocity
$\omega$. The local plasma pressure is positive on the outer surface
and negative at the inner surface; the numerical value of the
pressure in each case precisely balances the surface tensions at
these boundaries.

\begin{figure}
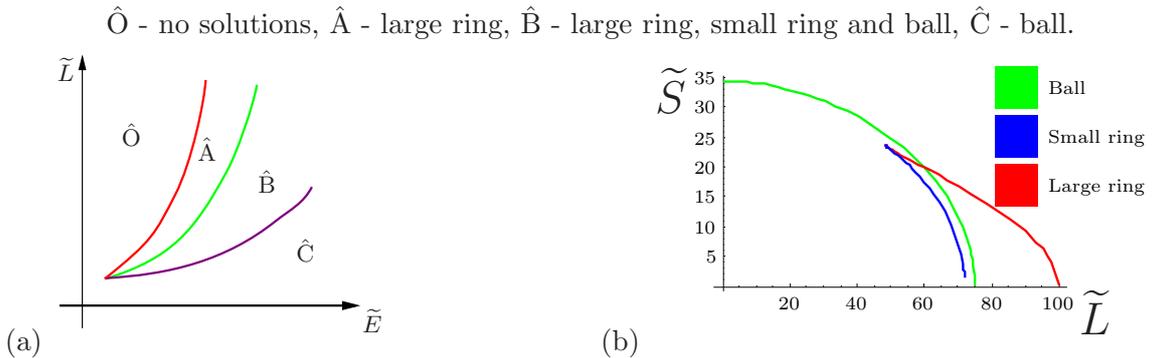

 \begin{center}
  \small{$\tOr$ - no solutions, $\tA$ - large ring, $\tB$ - large ring, small ring
  and ball, $\tC$ - ball.}\\
   \small{(a)}
   \input{allreg.tpx}
   \hspace{2.5cm}
   \small{(b)}
   \input{ent.tpx}
 \caption{(a) Regions where ball and ring solutions
exist, (b) their entropy as a function of angular momentum at fixed
energy.}\label{exist_intro:fig}
 \end{center}
\end{figure}

We now describe the moduli space of spinning plasmaball and plasma
ring solutions in a little more detail. In
fig.\ref{exist_intro:fig}(a) we have plotted the energy-angular
momentum plane, which we have divided up into 4 regions. In region
$\tC$ (low angular momentum at fixed energy) the only rigidly
rotating solution to the equations of fluid dynamics is the rotating
plasmaball. At higher angular momentum (region $\tB$) in addition to
the rotating plasmaball there exist two new annulus type solutions
which we call large and small ring solutions. As their names makes
clear, the solutions are distinguished by their size; the large ring
has a larger outer radius than the small one. On further raising
angular momentum (region $\tA$), the small ring and the ball cease
to exist; in this region the large ring is the only solution.
Finally, at still larger angular momentum (region $\tOr$) there
exist no solutions.

In fig.\ref{exist_intro:fig}(b) we have plotted the entropy of the
three different kinds of solutions as a function of their angular
momentum at a particular fixed energy. At angular momenta for which
all three solutions coexist (region $\tB$) the entropy of the small
ring is always smaller than the entropy of either the large ring or
the black hole. Upon raising the angular momentum, the solution with
dominant entropy switches from being the ball to the large ring; the
first order transition between these solutions occurs at an angular
momentum that lies on a `phase transition line' in the bulk of
region $\tB$. This picture suggests - and we conjecture - that the
ball and the large ring are locally stable with respect to
axisymmetric fluctuations, while the small ring is locally unstable
to such fluctuations.\footnote{It is possible that the large ring
exhibits Plateau-Rayleigh type instabilities that break rotational
invariance; such modes would map to Gregory-Laflamme type
instabilities of the bulk solution (see also \cite{Cardoso:2006ks}).
We thank T. Wiseman for suggesting this possibility.} In
\S\S\ref{sec:turn} we perform a `turning point' analysis of our
solutions, to find some evidence for this guess.

Let us now turn to the bulk dual interpretation of our solutions.
The fluid for the spinning  plasmaball is topologically a disk;
consequently the horizon topology for the dual bulk solution - the
$S^1$ fibration over this disk - yields an $S^3$. The bulk dual of
the spinning plasmaball is simply a rotating five dimensional black
hole. On the other hand the fluid configuration of the plasmaring
has the topology of $S^1\times$ interval; the $S^1$ fibration over
this configuration yields $S^1 \times S^2$; i.e. a five dimensional
black ring! Notice that in addition to the isometry along the $S^1$,
these ring solutions all have a isometry on the $S^2$ corresponding to translations
along the Scherk-Schwarz circle. This additional isometry, that does not appear
to be required by symmetry considerations, appears to be a feature of all known black ring solutions in flat space as well.

Using the gauge theory / gravity duality, the quantitative versions
of the fig.\ref{exist_intro:fig} give precise quantitative
predictions for the existence, thermodynamic properties and
stability of sufficiently big  black holes and black rings in
Scherk-Schwarz compactified AdS$_5$ spaces. While these
gravitational solutions have not yet been constructed, their
analogues in flat 5 dimensional space are known, and have been well
studied. The general qualitative features (and some quantitative
features) of fig.\ref{exist_intro:fig} are in remarkably good
agreement with the analogous plots for black holes and black rings
in flat five dimensional space (see \S\S\ref{sec:existcomp} for a
detailed discussion).

\begin{figure}
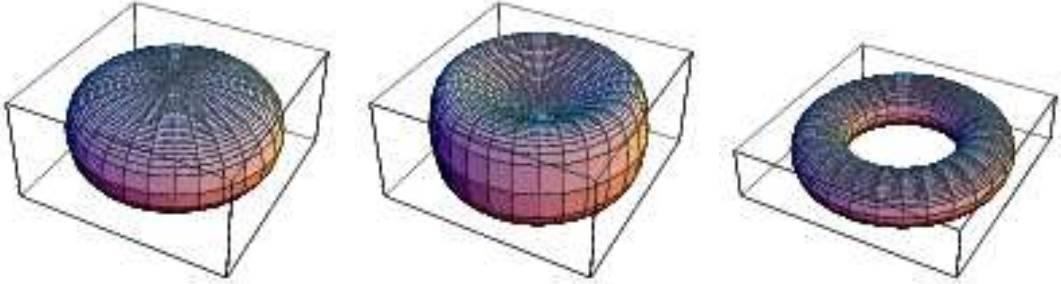

\begin{center}
  \input{obs.tpx}
  \input{pbs.tpx}
  \input{rs.tpx}
  \caption{Spinning ball and ring solutions.}\label{4d:fig}
\end{center}
\end{figure}

The constructions we have described above admit simple
generalisations to plasma solutions dual to black holes and black
rings in Scherk-Schwarz compactified AdS$_6$ space.\footnote{Note
that the spinning plasmaring has no analogue in 1+1 dimensional
fluid dynamics, for the excellent reason that there is no spin. This
tallies with the fact that there are no black rings in four
dimensions (at least in flat space).} As the qualitative nature of
the moduli space of black hole like solutions in six dimensional
gravity is poorly understood, this study is of interest. The
boundary duals of these objects, in the long wavelength limit, are
stationary solutions to the equations of fluid dynamics of a 4
dimensional field theory. In \S\ref{sec:4dim} we construct such
solutions. It turns out that these solutions occur in two
qualitatively distinct classes. The simplest solutions are simply
spinning balls of plasma; the fact that these balls spin causes them
to flatten out near the `poles'. As these balls are spun up, their
profile begins to `dip' near the poles (see fig.\ref{4d:fig}). As
these balls are further spun up, they pinch off at the centre and
turn into doughnut shaped rings (see fig.\ref{4d:fig}).

As in the three dimensional case, the horizon topology of the black
objects dual to the rotating plasmaballs and plasmarings described
above, is obtained by fibering the fluid configuration with an $S^1$
that shrinks to zero at the fluid edges. This procedure yields a
horizon topology $S^4$ for the dual to the rotating plasmaball, and
topology $S^3\times S^1$ for the dual to the plasmaring. As
plasmaball and plasmaring configurations appear to exhaust the set
of stationary fluid solutions to the equations of fluid dynamics, it
follows that arbitrarily large stationary black objects in
Scherk-Schwarz compactified AdS$_6$ all have one of these two
horizon topologies. $S^2 \times S^2$ is an example of another
topology one could have imagined for black objects in this space;
these would have been dual to hollow shells of rotating fluid;
however, there are no such stationary solutions to the equations of
fluid dynamics.

The analysis of four dimensional fluid configurations, described above,
demonstrates the power of the fluid dynamical method. In simple
contexts, the Navier-Stokes equations are much easier to solve than
the full set of Einstein's equations, and rather easily reveal
interesting and nontrivial information. It would be interesting to
extend our analysis of fluid dynamical models in various directions
to obtain information about the moduli space and stability of
classes of black solutions in AdS spaces. An obvious extension would
be to move to higher dimensions. As a first step in this direction,
we have obtained and partially solved the fluid flow equations in 5
dimensional spaces. A complete analysis of these equations would
yield the spectrum of black holes in Scherk-Schwarz compactified
AdS$_7$ spaces, in terms of the fluid dynamics of the deconfined
phase of the M5 brane theory on a Scherk-Schwarz circle.

Finally, we should point out that there has been a long history
within the General Relativity literature of treating black hole
horizons as surfaces associated with fluids. In one of the most
recent discussion within this framework, the authors of
\cite{Cardoso:2007ka} have modelled spinning black holes in $d+1$
dimensional (flat space) gravity by $d+1$ dimensional lumps of
incompressible fluid; here the fluid surface represents the black
hole horizon. Within this framework the 4+1 dimensional black ring,
for instance, is modelled by a 4+1 dimensional stationary fluid lump
of topology $B^3\times S^1$ \cite{Cardoso:2006sj}. This description
is rather different from the AdS/CFT induced description of black
rings in Scherk Schwarz compactified AdS$_5$ as a 2+1 dimensional
annulus of fluid. It would be interesting to better understand the
interconnections between these approaches.

\section{Fluid mechanics and thermodynamics}\label{sec:stress}

In this paper we study aspects of the dynamics of the deconfined
plasma described in the previous section. A full accounting for the
dynamics of the `gluon plasma' is a very complicated problem.
However, when the thermodynamic potentials and velocities vary over
length scales large compared to the quasiparticle (`gluon') mean
free path admit an effective description in terms of the equations
of fluid dynamics. The variables in this description are the local
values of the plasma or fluid velocity $u^\mu(x)$ and the plasma
density $\rho(x)$ (the equation of state, as discussed in
\S\S\ref{app:eqstate}, may be used to trade the density for the pressure
or the temperature). The equations of fluid dynamics are simply a
statement of the conservation of the stress tensor
\begin{equation}\label{Epconsv:eq}
  \nabla_\mu T^{\mu\nu} = \p_\mu T^{\mu\nu}
                        + \Gamma^\mu_{\mu\lambda} T^{\lambda\nu}
                        + \Gamma^\nu_{\mu\lambda} T^{\mu\lambda}
                        = 0\,.
\end{equation}
All input of the dynamical nature of the fluid that undergoes this
flow appears in the specification of the stress tensor in terms of
the velocity and density of the fluid, and from the thermodynamic
equation of state (which determines the pressure and temperature as
a function of density). In the rest of this brief subsection we
describe the functional form of the fluid dynamical stress tensor
for our plasma fluid  in detail.

The stress tensor can be split into three parts:
\begin{equation*}
  T^{\mu\nu} = T^{\mu\nu}_\mathrm{perfect} +
          T^{\mu\nu}_\mathrm{dissipative} +
          T^{\mu\nu}_\mathrm{surface}.
\end{equation*}
The first part, $T^{\mu\nu}_\mathrm{perfect}$, is the stress tensor
for a perfect fluid with no dissipative forces. It is a function
only of fluid velocity and thermodynamic quantities in the rest
frame, and not of their space time derivatives.

The second part, $T^{\mu\nu}_\mathrm{dissipative}$, receives
contributions from viscosity and heat flow. In the long wavelength
limit this piece is linear in the first derivatives of the velocity
and temperature.

The third part, $T^{\mu\nu}_\mathrm{surface}$, represents surface
contributions to the stress tensor, and requires more explanation.
Any fluid configuration with a surface has large variations in (for
instance) the fluid density over the scale of the mean free path, in
directions normal to the surface (see for instance
\cite{Aharony:2005bm}). As a consequence it is impermissable to use
the Navier-Stokes equations for the fluid in the neighbourhood of
the surface. When the deviations of the surface from a straight line
are small over length scale of the mean free path, however, all
effects of the surface may approximately be captured by a delta
function localised `surface tension' contribution,
$T^{\mu\nu}_\mathrm{surface}$, to the stress tensor. In the long
wavelength approximation, this term depends only on the gradients of
the surface and not its curvature.

\subsection{Perfect fluid stress tensor}\label{sec:blkstr}


The most general ultralocal stress tensor one can build out of the
fluid velocity and thermodynamic quantities that reduces to what is
expected for a fluid at rest is
\cite[ch.22]{Misner+ThorneETAL-Grav:73}
\begin{equation}\label{blkstr:eq}
  T^{\mu\nu}_\mathrm{perfect} = (\rho+P) u^\mu u^\nu + P g^{\mu\nu}\,.
\end{equation}

\subsection{Dissipative part}\label{sec:visc}

Realistic fluids have a dissipative component to their stress tensor
in addition to the perfect fluid piece. In the long wavelength
limit, this stress tensor is a function of  the acceleration,
expansion, projection, and shear tensors (see e.g. \cite[Exercise
22.6-7]{Misner+ThorneETAL-Grav:73} and references therein),
%
\begin{equation}\label{fluidtensors:eq}
\begin{split}
  a^\mu &= u^\nu \nabla_\nu u^\mu, \\
  \theta &= \nabla_\mu u^\mu, \\
  P^{\mu\nu} &= g^{\mu\nu} + u^\nu u^\mu, \\
  \sigma^{\mu\nu} &= \half \prn{P^{\mu\lambda} \nabla_\lambda u^\nu
                   + P^{\nu\lambda} \nabla_\lambda u^\mu}
                   - \frac{1}{d-1} \theta P^{\mu\nu}.
\end{split}
\end{equation}
In terms of these quantities and the heat flow vector $q^\mu$ (see
immediately below)
\begin{equation}\label{extraTvisc:eq}
  T^{\mu\nu}_\mathrm{dissipative} = -\zeta \theta P^{\mu\nu} -
  2\eta\sigma^{\mu\nu} + q^\mu u^\nu + u^\mu q^\nu\,,
\end{equation}
where $\zeta$ is the bulk viscosity, $\eta$ is the shear viscosity.
The heat flux vector,
\begin{equation}\label{heatcond:eq}
  q^\mu = -\kappa P^{\mu\nu} (\p_\nu\tloc + a_\nu\tloc)\,.
\end{equation}
is the  relativistic generalisation of $\vec{q} = -\kappa
\vec{\nabla} \tloc$ (here $\kappa$ is the thermal conductivity); the
extra term in \eqref{heatcond:eq} is related to the inertia of
flowing heat.

\subsection{Surface contribution}\label{sec:srfstr}

We will use a simple model of surface tension where the energy
stored in the surface and the force per unit length are both given
by $\sigma$, which we take to be the surface tension at the critical
temperature computed in \cite{Aharony:2005bm}. We will ignore any
dependance $\sigma$ could have on the fluid temperature. This
approximation is valid when the fluid temperature at the surface
does not deviate substantially from $\tc$.

Consider a localised lump of fluid whose surface in space is given
by the equation $f(x)=0$. The surface contribution to the stress
tensor will be proportional to $\sigma\delta(f)$. In the long
wavelength limit it will only depend on the first derivatives of
$f$. The most general stress tensor we can build is
\begin{equation*}
  T^{\mu\nu}_\mathrm{surface} = \brk{\alpha\, \p^\mu\! f\, \p^\nu\! f
                   +\beta\, u^\mu u^\nu
                   +\gamma\, (u^\mu\p^\nu\! f + \p^\mu\! fu^\nu)
                   +\delta\, g^{\mu\nu}}\sigma\delta(f)\,.
\end{equation*}
As $u^2=-1$ and $u^\mu\p_\mu f = 0$ (the surface moves with the
fluid), the only invariant quantity that
$\alpha,\beta,\gamma,\delta$ can depend on is $(\p^\mu\! f\, \p_\mu
f)$. We can fix this dependence by demanding invariance under
reparameterisations of the surface\footnote{However, we should
choose a parametrisation such that $\p_\mu f$ is well behaved at the
surface, e.g.\ $f=x$, but not $f=x^2$ or $f=\sqrt{x}$.} (e.g.
$f(x)\ra g(x)f(x)$, so that $\p f\ra g\p f + f\p g = g\p f$ at the
surface). Defining $f_\mu = \frac{\p_\mu f}{\sqrt{\p f\cdt\p f}}$:
\begin{equation*}
    T^{\mu\nu}_\mathrm{surface} = \brk{A f^\mu f^\nu
                   +B u^\mu u^\nu
                   +C \prn{u^\mu f^\nu + f^\mu u^\nu}
                   +D g^{\mu\nu}}\sigma\sqrt{\p f\cdt\p f}\delta(f)\,.
\end{equation*}
We can fix $A,B,C,D$ by looking at a fluid at rest, $u^\mu =
(1,0,0,\ldots)$, with a surface $f(x)=x$
\begin{equation*}
  T^{\mu\nu}_\mathrm{surface} =
  \begin{pmatrix}
    B-D & C   & 0  \\
    C   & A+D & 0  \\
    0   & 0   & D  \\
  \end{pmatrix}
     \sigma\delta(x)=
  \begin{pmatrix}
    1 & 0 &  0   \\
    0 & 0 &  0   \\
    0 & 0 & -1   \\
  \end{pmatrix}
     \sigma\delta(x).
\end{equation*}
This gives
\begin{equation}\label{srfStr:eq}
  T^{\mu\nu}_\mathrm{surface} = \sigma\brk{f^\mu f^\nu - g^{\mu\nu}}
               \sqrt{\p f\cdt\p f}\delta(f)\,.
\end{equation}

\subsection{Equations of state}\label{app:eqstate}

To solve the equations of fluid mechanics, one also needs expressions for
the various coefficients that appear in the stress tensor above in terms
of the density. For our purposes, we only need to know the thermodynamic
properties of the fluid, which could be determined from the static black
brane solution \eqref{AdSSchIb:eq}. In this subsection we discuss the free
energy, temperature etc.\ of
the plasma at rest. This is different from the free energy,
temperature etc.\ of the plasmaball/plasmaring. 

For a conformal theory in $d$ dimensions with no conserved charges,
dimensional analysis and extensivity determine
\begin{equation}\label{freecnfrm:eq}
  \floc = -\eos V \tloc^d,
\end{equation}
with $\eos$ an arbitrary constant. In our situation, the plasma
is dual to the same black brane, so it doesn't know about any capping off
in the IR except that the energy is measured with respect to a
different zero. Before reducing on the Scherk-Schwarz circle, it
behaves like a conformal theory in $d+1$ dimensions plus a vacuum
energy density. After dimensional reduction\footnote{Strictly
speaking, it is not a dimensional reduction as we will have plasma
temperature of the same order as the Kaluza-Klein scale. Rather, we
are restricting attention to classical solutions that do not vary in
this compact dimension.}, we have
\begin{equation}\label{freecnfng:eq}
  \floc = V\prn{\rz - \eos \tloc^{d+1}}.
\end{equation}
This gives
\begin{equation}\label{therm:eq}
  \begin{split}
    P &= -\pdiffc[\tloc]{\floc}{V} = \eos \tloc^{d+1} - \rz \,,\\
    \sloc &= -\pdiffc[V]{\floc}{\tloc} = (d+1)\eos V \tloc^{d} \,,\\
    \eloc &= \floc+\tloc\sloc = V\prn{\rz + d\eos \tloc^{d+1}}\,.
  \end{split}
\end{equation}
In terms of intensive quantities, we have
\begin{equation}\label{thermint:eq}
  \begin{aligned}
    P &= \frac{\rho-(d+1)\rz}{d} \,,&\qquad
    P+\rho &= \prn{\frac{d+1}{d}}(\rho-\rz) \,,\\
    s  &=
    (d+1)\eos^{1/(d+1)}\prn{\frac{\rho-\rz}{d}}^{d/(d+1)},&
    \tloc &= \prn{\frac{\rho-\rz}{d\eos}}^{1/(d+1)},
  \end{aligned}
\end{equation}
or, in three dimensions
\begin{equation}\label{therm3d:eq}
  \begin{aligned}
    P &= \frac{\rho-4\rz}{3} \,,&\qquad
    P+\rho &= \frac{4}{3}(\rho-\rz) \,,\\
    s  &=
    \frac{4\eos^{1/4}}{3^{3/4}}\prn{\rho-\rz}^{3/4},&
    \tloc &= \prn{\frac{\rho-\rz}{3\eos}}^{1/4}.
  \end{aligned}
\end{equation}

Note that the critical density and temperature are those for which
the pressure is zero
\begin{equation}\label{crittherm:eq}
    \rho_c = (d+1)\rz\,, \qquad
    \tc = \prn{\frac{\rz}{\eos}}^{1/(d+1)}.
\end{equation}
For the black-brane equation of state \eqref{equationofstate:eq}
\begin{equation}\label{braneparam:eq}
  \rz = \frac{\pi^2 N^2 \tc^3}{8}\,,
  \qquad
  \eos = \frac{\pi^2 N^2}{8 \tc}\,.
\end{equation}
However, the values of these constants will not be important below.

\section{Rigidly rotating configurations}\label{sec:eqmot}

In this section, we study stationary, axially symmetric rotating
fluid configurations, whose equation of state is  presented in
various forms in \S\S\ref{app:eqstate}. We choose the axis of
rotation as our origin in polar coordinates; in these coordinates
the fluid density is a function only of the radial coordinate $r$,
and the $(t,r,\phi)$ components of the velocity are given by $u^\mu
= \gamma(1,0,\omega)$ with $\gamma=\prn{1-\omega^2r^2}^{-1/2}$.  We
will find two distinct kinds of solutions; rotating plasmaballs with
the topology of a two dimensional disk, and plasmarings with the
topology of a two dimensional annulus. The configurations we find
are exact solutions to the equations of relativistic fluid dynamics;
in \S\S\ref{sec:validity} we will demonstrate that these equations
accurately represent plasma dynamics for large enough plasmaballs
and plasmarings.

\subsection{Equations of motion}

Our fluid propagates in flat 2+1 dimensional space. In polar
coordinates
\begin{equation}\label{metric:eq}
  \dr s^2 = -\dr t^2 + \dr r^2 + r^2 \dr \phi^2\,.
\end{equation}
This gives the following non-zero Christoffel symbols:
\begin{equation}\label{chrst:eq}
  \Gamma^r_{\phi\phi} = -r \qquad
  \Gamma^\phi_{r\phi} = \Gamma^\phi_{\phi r} =
  \frac{1}{r}\,.
\end{equation}

For the  stationary, axially symmetric configurations under
consideration, $\p_t T^{\mu \nu}= \p_\phi T^{\mu \nu} = 0$. Using
\eqref{chrst:eq}, \eqref{Epconsv:eq} becomes
\begin{align}\label{tEpconsv:eq}
    0 &= \nabla_\mu T^{\mu t}
       = \p_r T^{rt} + \frac{1}{r}T^{rt}, \\ \label{rEpconsv:eq}
    0 &= \nabla_\mu T^{\mu r}
       = \p_r T^{rr} + \frac{1}{r}T^{rr} - r T^{\phi\phi}, \\ \label{fEpconsv:eq}
    0 &= \nabla_\mu T^{\mu \phi}
       = \p_r T^{r\phi} + \frac{3}{r}T^{r\phi}.
\end{align}

The boundaries are $f_n = r-r_n$, with $n$ labelling the different
boundaries (outer for the disk, outer and inner for the annulus).
The `perfect fluid part' of the stress tensor is
\begin{equation}\label{blkstrour:eq}
  T^{\mu\nu}_\mathrm{perfect} =
   \begin{pmatrix}
     \gamma^2(\rho+\omega^2r^2P) & 0 & \gamma^2\omega(\rho+P) \\
     0                           & P & 0 \\
     \gamma^2\omega(\rho+P)      & 0 & \frac{\gamma^2}{r^2}(\omega^2r^2\rho+P) \\
   \end{pmatrix}
\end{equation}
and the surface stress tensor
\begin{equation}\label{srfstrour:eq}
  T^{\mu\nu}_\mathrm{surface} =
    \sigma \sum_n \delta(r-r_n)
  \begin{pmatrix}
    1 & 0 & 0 \\
    0 & 0 & 0 \\
    0 & 0 & -\frac{1}{r^2} \\
  \end{pmatrix}
\end{equation}
For the dissipative part of the stress tensor, we find
$\theta=\sigma^{\mu\nu}=0$ and
%
%
\begin{equation}\label{ourheatcond:eq}
  \p_\nu\tloc + a_\nu\tloc = \prn{0, \gamma
  \diff{}{r}\brk{\frac{\tloc}{\gamma}},0}\,,
\end{equation}
so that
\begin{equation}\label{dissT:eq}
  T^{\mu\nu}_\mathrm{dissipative} =
   -\kappa\gamma^2\diff{}{r}\brk{\frac{\tloc}{\gamma}}
   \begin{pmatrix}
     0 & 1      & 0      \\
     1 & 0      & \omega \\
     0 & \omega & 0      \\
   \end{pmatrix}
\end{equation}

We will now write the equations of motion $\nabla_\mu T^{\mu \nu}=0$
temporarily ignoring the contribution from  this heat flow,
$T^{\mu\nu}_\mathrm{dissipative}$; it will turn out (we see this
immediately below) that $T^{\mu\nu}_\mathrm{dissipative}$ actually
vanishes on our solutions, justifying this procedure.

The only non-trivial equation of motion, \eqref{rEpconsv:eq}, can be
written as
\begin{equation}\label{eom:eq}
  \diff{P}{r} = \frac{\omega^2r}{1-\omega^2r^2}\,(\rho+P)
              - \sum_n \frac{\sigma}{r}\, \delta(r-r_n)\,.
\end{equation}
For these fluids (with no chemical potentials for any conserved
charges), $P=-f$ is a function only of $\tloc$, so $P+\rho = s\tloc$
and $\diff{P}{\tloc} = s$. So, away from the boundaries,
\eqref{eom:eq} becomes
\begin{equation}\label{tempr:eq}
\begin{split}
  s \diff{\tloc}{r} &= s \tloc \diff{\ln\gamma}{r} \\ \implies
  \diff{}{r}\brk{\frac{\tloc}{\gamma}} &= 0
\end{split}
\end{equation}
It follows that $T^{\mu \nu}_\mathrm{dissipative}$  vanishes for
rigid rotation, justifying our neglect of heat flow.

Our discussion has not assumed a specific form of the equation of
state. Using this particular equation of state of our plasma
\eqref{therm3d:eq}, we can rewrite \eqref{tempr:eq} in the fluid
interior as
\begin{equation}\label{eombulk:eq}
  \prn{\rho(r)-\rz}\prn{1-\omega^2r^2}^2 = \text{constant}.
\end{equation}
Integrating \eqref{eom:eq} across a surface gives
\begin{equation}\label{bcndgen:eq}
  P_>-P_< = -\frac{\sigma}{r}\,.
\end{equation}
where $P_>$ and $P_<$ are the pressures at infinitesimally greater
and smaller radii than the location of the surface.

\subsection{Spinning ball}\label{sec:ballsol}

Let us first study a fluid configuration with a single outer surface
at $r=\ro$ with $P_>=0$. Using the equation of state
\eqref{therm3d:eq}, the boundary condition \eqref{bcndgen:eq} can be
written as
\begin{equation}\label{bcout:eq}
  \rho(\ro) = 4\rz + \frac{3\sigma}{\ro}\,.
\end{equation}

If we define dimensionless variables
\begin{equation}\label{newvars:eq}
    \tw = \frac{\sigma\omega}{\rz} \,,   \qquad
    \z = \frac{\rz r}{\sigma} \,,   \qquad
    v = \omega r = \tw \z \,,
\end{equation}
then \eqref{eombulk:eq} can be written as
\begin{equation}\label{ballsol:eq}
  \prn{\frac{\rho(v)-\rz}{3\rz}}\prn{1-v^2}^{2}
       = \prn{1+\frac{\tw}{\vo}}\prn{1-\vo^2}^{2}
       \equiv g_+(\vo).
\end{equation}

Note that the range of $v$ is $[0,1]$ and $\rho(v)-\rz$ is always
positive for this solution, as is required for the last equation of
\eqref{therm3d:eq} to make sense.

We can also compute the local plasma temperature using
\eqref{therm3d:eq}
\begin{equation}\label{temploc:eq}
  \tloc = \gamma\prn{\frac{\rz g_+(\vo)}{\eos}}^{1/4}.
\end{equation}

\subsection{Spinning ring}\label{sec:ringsol}

We now turn to solutions that have an inner surface well an outer
surface. In addition to the boundary condition at the outer radius
\eqref{bcout:eq} we now have
\begin{equation}\label{bcin:eq}
  \rho(\ri) = 4\rz - \frac{3\sigma}{\ri}\,.
\end{equation}

So the full solution is
\begin{equation}\label{ringsol:eq}
 \begin{split}
  \prn{\frac{\rho(v)-\rz}{3\rz}}\prn{1-v^2}^{2}
       = \prn{1+\frac{\tw}{\vo}}\prn{1-\vo^2}^{2}
       &\equiv g_+(\vo)\\
       = \prn{1-\frac{\tw}{\vi}}\prn{1-\vi^2}^{2}
       &\equiv g_-(\vi).
 \end{split}
\end{equation}

Note that $\rho(v)-\rz\geq 0$ provided that $\vi \geq \tw$.


Again, we can compute the local plasma temperature using
\eqref{therm3d:eq}
\begin{equation}\label{temploc2:eq}
  \tloc = \gamma\prn{\frac{\rz g_+(\vo)}{\eos}}^{1/4}
        = \gamma\prn{\frac{\rz g_-(\vi)}{\eos}}^{1/4}.
\end{equation}

\begin{figure}
  \begin{center}
    \input{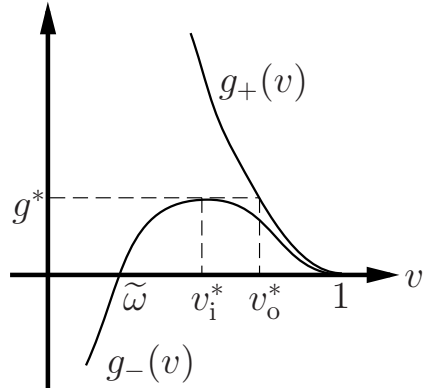}
  \caption{Graph of $g_\pm(v)$ showing possible values of $v_{o,i}$}\label{gpm:fig}
  \end{center}
\end{figure}

The two functions, $g_\pm(v)$ are schematically plotted in
fig.\ref{gpm:fig} for some value of $\tw$, where we have labelled
special velocities $\vi^*$ and $\vo^*$. As $\vi<1$, it is necessary
that $\tw<1$.

We can see that there are no solutions to $g_+(\vo)=g_-(\vi)$ for
$\vo<\vo^*$ and two solutions for $\vo>\vo^*$. One of these has
$\vi<\vi^*$ (the thick ring) and one has $\vi>\vi^*$ (the thin
ring). The distinction between `thin' and `thick' rings will not
prove physically important. In \S\S\ref{sec:exist} we will find it
physically useful to distinguish between distinct ring solutions (we
will call these large and small rings) at the same values of
conserved charges (energy and angular momentum), rather than the
parameters $\vo$ and $\tw$.

\section{Thermodynamic Potentials}\label{sec:charges}

In this section, we compute the thermodynamic potentials (energy,
angular momentum, entropy, etc.) for the spinning plasmaball and
plasmarings themselves, rather than their constituent plasma. This
includes contributions from the kinetic energy of the plasma as well
as its internal energy.

The constitutive relations we find  are predictions for, e.g.,
entropy as a function of mass and angular momentum of the dual
gravity solutions.

\subsection{Densities}\label{sec:chdens}

In this subsection we list formulae for energy density, angular
momentum density and entropy density. In the next subsection we will
integrate these expressions to find explicit formulae for the
energy, angular momentum and entropy of spinning plasmaballs and
plasmarings.

The energy density is given by
\begin{equation}\label{Edens:eq}
\begin{split}
  T^{tt} &= \gamma^2 \prn{\rho+\omega^2r^2P}
              + \sum_n \sigma\, \delta(r-r_n) \\
     &= \rz \brk{1+g_+(\vo)\frac{3+v^2}{\prn{1-v^2}^{3}}
                + \sum_n 2\tw v \,\delta(v^2-v^2_n)}.
\end{split}
\end{equation}

The angular momentum density is given by
\begin{equation}\label{Ldens:eq}
  r^2 T^{t\phi} = \gamma^2\omega r^2 (\rho+P) =
    4\sigma\frac{g_+(\vo)}{\tw}\frac{v^2}{\prn{1-v^2}^{3}}\,.
\end{equation}

The entropy density is given by
\begin{equation}\label{Sdens:eq}
  \gamma s =
  \frac{4\eos^{1/4}}{3^{3/4}}\frac{(\rho-\rz)^{3/4}}{\sqrt{1-v^2}} =
  4(\eos\rz^3)^{1/4}\frac{g_+(\vo)^{3/4}}{(1-v^2)^{2}}
\end{equation}

\subsection{Integrals}\label{sec:ints}
%

We can define some dimensionless variables
\begin{equation}\label{redch:eq}
  \tE = \frac{\rz E}{\pi\sigma^2} \,,\quad
  \tL = \frac{\rz^2 L}{\pi\sigma^3} \,,\quad
  \tS = \frac{\rz^{5/4}S}{\pi\eos^{1/4}\sigma^2} \,,\quad
  \tT = T\prn{\frac{\eos}{\rz}}^{1/4},\quad
  \tO = \frac{\sigma\Omega}{\rz}\,.
\end{equation}
The last of these ensure that $\tT = \pdiffc[\tL]{\tE}{\tS}$ and
$\tO = \pdiffc[\tS]{\tE}{\tL}$ follow from $T = \pdiffc[L]{E}{S}$
and $\Omega = \pdiffc[S]{E}{L}$

\subsubsection{Spinning ball}\label{sec:ballch}

Energy
\begin{equation}\label{ballE:eq}
  \tE = \frac{4\vo^2-\vo^4+5\tw \vo-\tw \vo^3}{\tw^2}
\end{equation}
Angular momentum
\begin{equation}\label{ballL:eq}
  \tL = \frac{2\vo^4+2\tw \vo^3}{\tw^3}
\end{equation}
Entropy
\begin{equation}\label{ballS:eq}
  \tS = \frac{4\vo^2}{\tw^2}\sqrt{1-\vo^2}
         \prn{1+\frac{\tw}{\vo}}^{3/4}
\end{equation}

\subsubsection{Spinning ring}\label{sec:ringch}

Energy
\begin{equation}\label{ringE:eq}
\begin{split}
    \tE &= \frac{4(\vo^2-\vi^2) - (\vo^4-\vi^4)
                 + 5\tw (\vo+\vi) - \tw (\vo^3+\vi^3)}
             {\tw^2}
\end{split}
\end{equation}
Angular momentum
\begin{equation}\label{ringL:eq}
\begin{split}
  \tL &= \frac{2(\vo^4-\vi^4)+2\tw (\vo^3+\vi^3)}{\tw^3}
\end{split}
\end{equation}
Entropy
\begin{equation}\label{ringS:eq}
\begin{split}
  \tS &= \frac{4}{\tw^2}
        \brk{\sqrt{1-\vo^2}\prn{1+\frac{\tw}{\vo}}^{3/4}
            -\sqrt{1-\vi^2}\prn{1-\frac{\tw}{\vi}}^{3/4}}
\end{split}
\end{equation}

\subsection{Temperature and angular velocity}\label{sec:temp}

In this subsection we determine the temperature and angular velocity
of spinning plasmaballs and plasmarings using
\begin{equation}\label{tempdef:eq}
  \tT = \pdiffc[\tL]{\tE}{\tS}, \qquad   \tO = \pdiffc[\tS]{\tE}{\tL}.
\end{equation}

Note that the temperature defined above is different from the local
plasma temperature, $\tloc$ (which varies across our solutions), in
(\ref{therm:eq}-\ref{therm3d:eq}); the angular velocity defined
above will turn out to be $\omega$ on all our solutions, although it
is apparently a priori different.

It may be verified that the temperature and angular velocity of
plasmaballs is given by
\begin{equation}\label{temp:eq}
  \tT = [g_+(\vo)]^{1/4}, \qquad \tO = \tw\,.
\end{equation}
The corresponding expressions  for the rings are identical
\begin{equation}\label{temp2:eq}
  \tT = [g_+(\vo)]^{1/4} = [g_-(\vi)]^{1/4}, \qquad \tO = \tw\,.
\end{equation}

Thus, local temperatures, $\tc$, and angular velocities, $\omega$,
for both the ball and the ring, are given simply in terms of $T$ and
$\Omega$
\begin{equation*}
  \tloc = \frac{T}{\sqrt{1-v^2}}\,, \qquad \omega=\Omega\,.
\end{equation*}

\section{Solutions at fixed energy and angular
momentum}\label{sec:enang}

\subsection{Existence}\label{sec:exist}

\begin{figure}
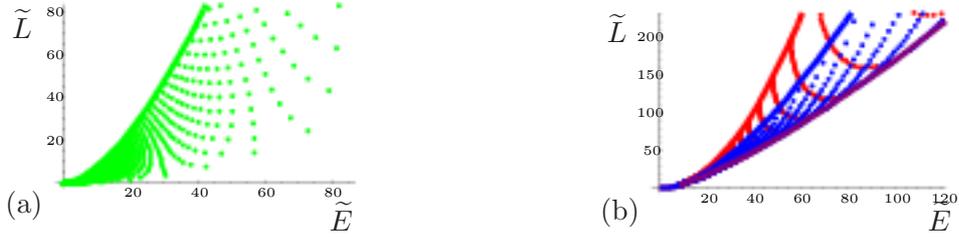

\begin{center}
  \input{ballexist.tpx}
  \hspace{2.5cm}
  \input{ringexist.tpx}
  \caption{Scatter plots of energy and angular momentum of (a)
spinning balls and (b) rings.}\label{exist_scat:fig}
\end{center}
\end{figure}

In fig.\ref{exist_scat:fig}, we display scatter plots for the energy
and angular momentum of ball and ring solutions over the full range
of solution parameters.\footnote{In order to generate these plots
for the ball, a range of vales of $(\vo,\tw)$ were chosen and
$(E,L)$ were computed using (\ref{ballE:eq},\ref{ballL:eq}). For the
ring, a range of vales of $(\vi,\tw)$ were chosen, $\vo$ was
computed using \eqref{ringsol:eq}, and $(E,L)$ were computed using
(\ref{ringE:eq},\ref{ringL:eq}).} The various regions of existence
of the plasmaball, thin plasmaring and thick plasmaring in the $E-L$
plane are drawn schematically in fig.\ref{exist_prelim:fig}.

\begin{figure}
 \begin{center}
 \small{O - no solutions, A - 1 ring, B - 2 rings, C - 1 ball.}\\
  (a)
  \input{ballreg.tpx}
  \hspace{2.5cm}
  (b)
  \input{ringreg.tpx}
 \caption{Regions where (a) ball and (b) ring solutions
exist.}\label{exist_prelim:fig}
 \end{center}
\end{figure}

The ball solution exists over a region C in the $E-L$ plane. At the
boundary of the region C the ball solution $\vo$ attains its maximum
value of unity. Using (\ref{ballE:eq},\ref{ballL:eq}) we find an
analytic expression for the boundary of C:
\begin{equation}\label{ballexist:eq}
  \tL = \frac{2}{27}\brk{(3\tE+4)^{3/2}-9\tE-8}
     \sim \frac{2\tE^{3/2}}{3^{3/2}}
     \quad\text{for large }\tE\,.
\end{equation}
From \eqref{temp:eq}, we see that balls on this boundary saturate
the extremality bound (i.e.\ have zero temperature).

Like the balls, rings of a fixed energy have a maximum value of
angular momentum. Rings at the edge of this bound (the boundary
between O and A in fig.\ref{exist_prelim:fig}b) have $\vo=\vi=1$ and
so are extremal (see \eqref{temp2:eq}) and of zero width. Using
(\ref{ringE:eq},\ref{ringL:eq}) the O-A boundary is given by
\begin{equation}\label{thinexist:eq}
  \tL = \frac{\tE^2}{16}\,,
\end{equation}
(this expression is valid only for $\tE>8$, $\tL>2$; at lower
energies $\tw$ exceeds unity).

As we lower angular momentum of the solution, this ring moves away
from extremality and increases in width. At a particular angular
momentum (the boundary between region A and region B) a new ring
solution comes into existence. The corresponding solution has
$\vo=1$, $\vi=\tw$ and so is extremal (see \eqref{temp2:eq}. Using
(\ref{ringE:eq},\ref{ringL:eq}), the analytic expression for the A-B
boundary is given by
\begin{equation}\label{thickexist:eq}
  \tL = \frac{2}{27}\brk{(3\tE+1)^{3/2}-9\tE+1}
     \sim \frac{2\tE^{3/2}}{3^{3/2}}
     \quad\text{for large }\tE\,,
\end{equation}
(for $\tE>8$, $\tL>2$ as above). In the high energy limit  $\tE \gg
1$ the ratio of angular momentum for the new extremal rings (at the
A-B boundary) and extremal plasmaball tends to unity, (even though
the the difference between angular momenta does not go to zero).
Consequently the leading high energy behaviour of
\eqref{thickexist:eq} and \eqref{ballexist:eq} is the same in this
limit,  as is also clear from fig.\ref{exist_final:fig}. We
emphasise that, for our solutions, the extremal ball and extremal
thick ring are not quite identical (as is the case for black holes
and small black rings \cite{Emparan:2001wn} in flat space) as the
inner radius of our extremal thick rings does not vanish. However,
the inner radius of the extremal thick ring is always (for all
values of energy) of the same order as the thickness of the domain
wall. As the fluid dynamics approximations fail precisely under
these conditions, it could well be that the new extremal plasmaring
and extremal plasmaball are actually identical configurations.

\begin{figure}
 \begin{center}
   \input{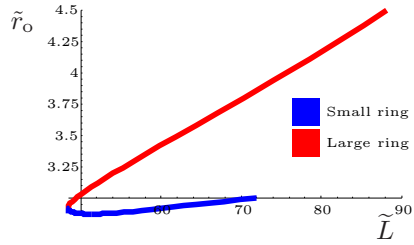}
 \caption{Outer radius of large and small rings as a function of
angular momentum, $\tL$, at fixed energy,
$\tE=40$.}\label{outer:fig}
 \end{center}
\end{figure}

\begin{figure}
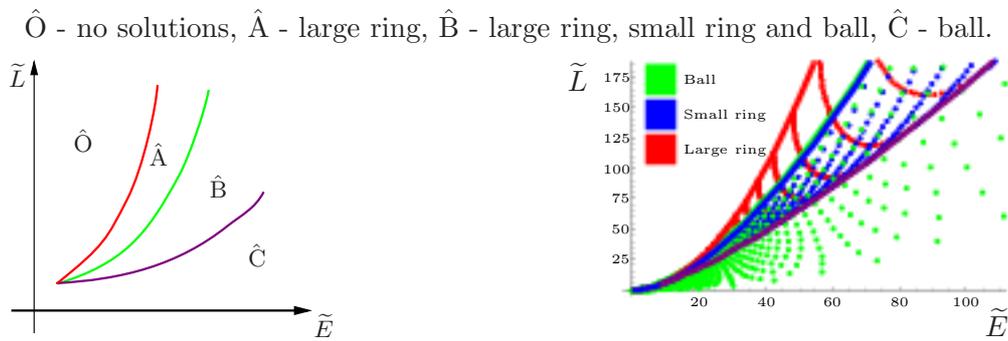

 \begin{center}
  \small{$\tOr$ - no solutions, $\tA$ - large ring, $\tB$ - large ring, small ring
  and ball, $\tC$ - ball.}\\
   \input{allreg.tpx}
   \hspace{2.5cm}
   \input{allexist2.tpx}
 \caption{Regions where ball and ring solutions exist.}
\label{exist_final:fig}
 \end{center}
\end{figure}

As we further lower angular momentum, the new ring solution moves
away from extremality; this new solution always has a smaller outer
radius than the `original' ring solution (the solutions that also
exists in region A), as shown in fig.\ref{outer:fig}. As a
consequence, we refer to these two ring solutions as small and large
respectively.

Further lowering angular momentum, we hit the boundary between
regions B and O where the two ring solutions merge into each other.
At still lower angular momentum, we have no ring solutions. The C-O
boundary may thus be obtained by minimising $\tL$ at fixed $\tE$. If
one uses \eqref{ringsol:eq} to eliminate $\tw$, this amounts to
\begin{equation*}
  \pdiffc[\vi]{\tE}{\vo}\pdiffc[\vo]{\tL}{\vi}
  -\pdiffc[\vo]{\tE}{\vi}\pdiffc[\vi]{\tL}{\vo}
    =0.
\end{equation*}

The existence of plasmaball and plasmaring solutions in the E-L
plane may thus be summarised as in fig.\ref{exist_final:fig}.

%

\subsection{Validity}\label{sec:validity}

As we have described above,  plasmaballs and plasmarings are exact
solutions to the relativistic Navier-Stokes equations (supplemented
by sharp surface boundary conditions). However these equations of
fluid dynamics accurately capture the dynamics of the fluid plasma
only under certain conditions. In our discussion we have assigned a
well defined pressure and temperature to the fluid at each point in
space. Clearly this procedure is valid only when the variation of
these thermodynamic quantities is small over the length scale of the
mean free path of the quasiparticles (roughly gluons) of our system.
The mean free path is of the same order as the mass gap of the
theory, which in turn is similar to the deconfinement
temperature.\footnote{It was argued in \cite{Aharony:2005bm} that
the mean free path does not scale with $N$. According to
quasiparticle kinetic theory, the mean free path is approximately
the ratio of the shear viscosity and the energy density. The
computations reviewed in \cite{Son:2007vk} show that this quantity
is of order $1/\tc$ in the limit of the 't~Hooft coupling
$\lambda\ra\infty$.}

Second, we have treated the surface of the plasma as sharp; in
reality this surface has a thickness of order $\tc^{-1}$.
Consequently, our treatment of the surface is valid only when its
deviation from a straight line occurs on scales large compared to
$\tc^{-1}$ (higher derivative contributions to the surface stress
tensor, which we have ignored in our treatment, would become
important if this were not the case); further we must also require
that only a small fraction of the fluid should reside in surfaces.

Thirdly, we have ignored the fact that the surface tension is a
function of the fluid temperature at the surface, and simply set
$\sigma=\sigma(\tc)$. This is valid provided that $\tloc/\tc \approx
1$ at all surfaces.

Finally, the fluid evolution equations, by their very nature, track mean
velocities and energy densities, ignoring fluctuations. In our context this
approximation is justified by large $N$; fluctuations are suppressed by powers of $1/N^2$,
dual to the suppression of quantum metric fluctuations in the bulk.

Recall that
\begin{equation*}
  \rz \sim N^2\tc^3, \qquad \sigma \sim  N^2\tc^2,
  \qquad \text{so}\quad \frac{1}{\tc} \sim \frac{\sigma}{\rz}
\end{equation*}
(for the domain wall solution of \cite{Aharony:2005bm},
$\frac{\sigma}{\rz}=2.0\times \frac{4}{\tc}$ and the thickness is
$6\times \frac{1}{2\pi \tc}$).

As an estimate of the scale over which thermodynamic quantities
vary, we compute the fractional change in the fluid temperature over
the distance of a mean free path. As the temperature is proportional
to $\gamma$, we should look at
\begin{equation*}
  \frac{1}{\tc}\diff{}{r}\ln\gamma \sim
   \frac{\sigma}{\rz} \frac{\omega^2 r}{1-\omega^2r^2} =
   \frac{\tw v}{1-v^2}\,.
\end{equation*}
As this takes its maximum value at the outer surface, the condition
for the validity of the equations of fluid dynamics  may be
estimated to be
\begin{equation}\label{validity2:eq}
  \Delta u \equiv \frac{\tw \vo}{1-\vo^2} \ll 1\,.
\end{equation}

Our treatment of the surface as a zero-thickness object is valid if
\begin{equation*}
  \{\ro,\ri,\ro-\ri\} \gg \frac{1}{\tc} \sim
  \frac{\sigma}{\rz}
\end{equation*}
(for the ring, the $\ro$ inequality in the equation above follows
automatically from the either of the other two inequalities). This
condition can be rewritten in terms of our dimensionless variables
as
\begin{equation}\label{validity:eq}
\begin{aligned}
  \vo &\gg \tw
      &\quad&\text{or}\quad&
   \zo &\gg 1
      &\quad \text{for the ball,}\\
  \{\vi,\vo-\vi\} &\gg \tw
      &\quad&\text{or}\quad&
   \{\zi,\zo-\zi\} &\gg 1
      &\quad \text{for the ring.}
\end{aligned}
\end{equation}

\begin{figure}[p]
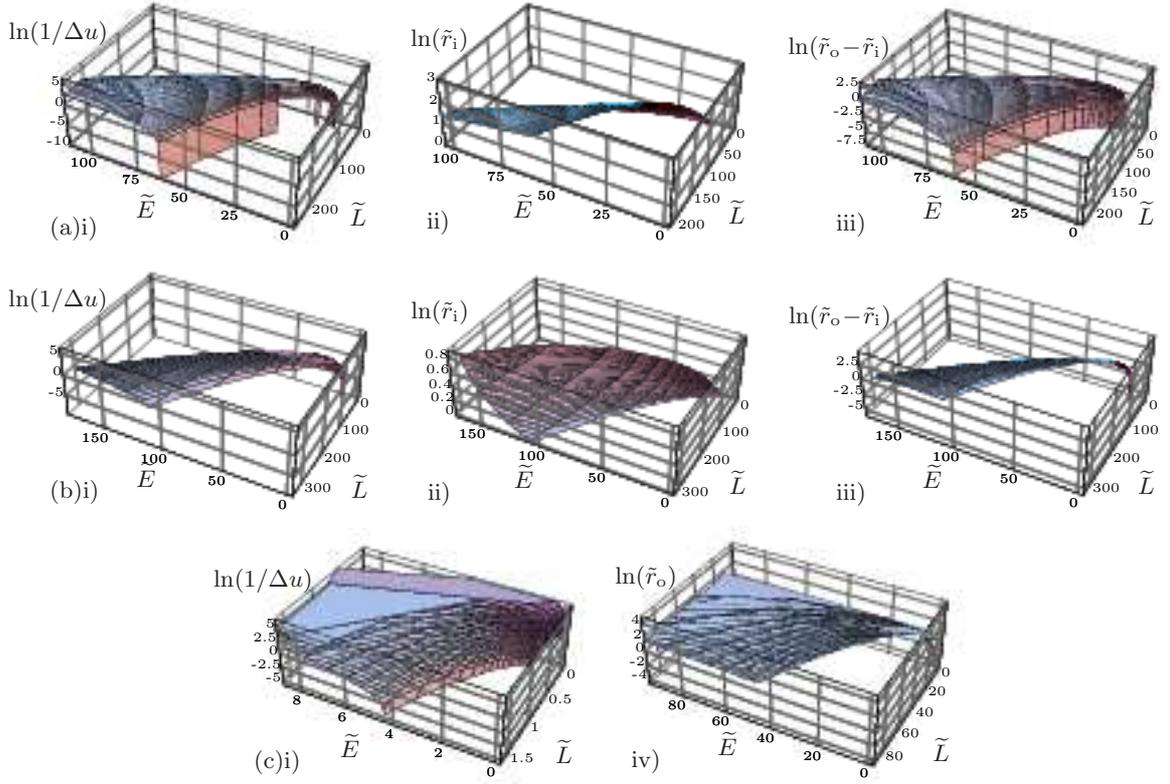

 \begin{center}
   \input{valthindu.tpx}
   \input{valthinzi.tpx}
   \input{valthinzozi.tpx}\\
   \input{valthickdu.tpx}
   \input{valthickzi.tpx}
   \input{valthickzozi.tpx}\\
   \input{valballdu.tpx}
   \input{valballzo.tpx}
 \end{center}
 \caption{Plots of i) $\ln(1/\Delta u)$ ii) $\ln(\zi)$, iii) $\ln(\zo-\zi)$,
iv) $\ln(\zo)$ for (a) large rings, (b) small rings and (c) balls.}
\label{validity:fig}
\end{figure}

In fig.\ref{validity:fig}, we have plotted $\ln(1/\Delta u)$,
$\ln(\zi)$, $\ln(\zo-\zi)$ and $\ln(\zo)$ for the thin ring, thick
ring and ball. From the figure we observe that these quantities are
large (and so the fluid dynamics approximations of this paper are
accurate) when our rings and balls have large energy and we stay
away from the extremality bounds.

\begin{figure}[p]
 \begin{center}
   \input{linvalballzero.tpx}
   \input{linvalthickvi.tpx}
   \input{linvalthinvi.tpx}\\
   \input{linvalballvo.tpx}
   \input{linvalthickvo.tpx}
   \input{linvalthinvo.tpx}
 \end{center}
 \caption{Plots of (a) $\ln(\tloc_{min}/\tc)$ (b)
$\ln(\tloc_{max}/\tc)$ for i) balls, ii) small rings, iii) large
rings.} \label{linvalidity:fig}
\end{figure}

Finally,  validity of our approximation of the surface tension as
a constant (independent of temperature) requires that the maximum
and minimum values of $\ln(\tloc/\tc)$ (which occur at the outer and
inner surfaces respectively) are both small. We have plotted these
quantities in fig.\ref{linvalidity:fig}. It is clear from these
figures that this condition is fulfilled for large energy and
angular momentum provided that we are not near extremality.


\subsection{Global stability and phase diagram}\label{sec:globstab}

\begin{figure}[tbh]
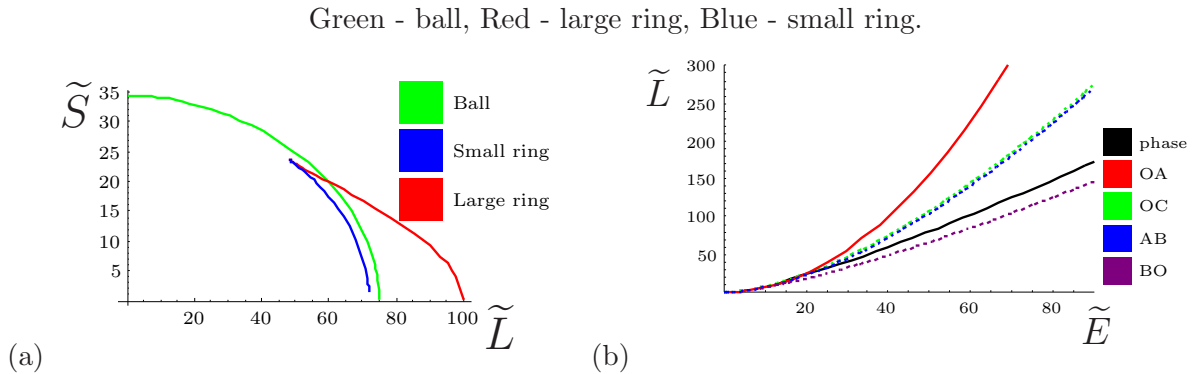

 \begin{center}
  \small{Green - ball, Red - large ring, Blue - small ring.} \\
  (a)\input{ent.tpx}
  (b)\input{bndph.tpx}
 \caption{(a) Entropy, $\tS$, as a function of angular momentum, $\tL$
for fixed energy, $\tE=40$. (b) Phase boundary with existence boundaries.}\label{ent:fig}
 \end{center}
\end{figure}

Recall that (see fig.\ref{exist_final:fig}) at fixed values of
energy and angular momentum, we have either 0, 1 or 3 plasmaball /
plasmaring solutions. At those values of charges for which multiple
solutions exist, it is natural to inquire which of these solutions
is entropically favoured. In fig.\ref{ent:fig}(a) we have plotted the
entropy of plasma ball and plasmaring solutions as a function of
angular momentum at fixed energy.

Note that, when it exists, the small ring always carries lower
entropy than both the big ring and the plasmaball. At low enough
angular momentum the plasmaball is the only solution. This solution
continues to be entropically dominant (upon raising the angular
momentum) over an interval, even after the new ring solutions are
nucleated. At a critical angular momentum, however, the entropy of
the large ring equals and then exceeds the entropy of the plasmaball
(all three ring solutions continue to exist in a neighbourhood about
this point). The large ring is the entropically dominant solution at
all larger angular momenta.

The phase boundary can be seen in fig.\ref{ent:fig}(b).


\subsection{Comparison with black rings in flat 5D space}\label{sec:existcomp}

As we have explained in the introduction, the plasmaball and
plasmaring solutions of this paper are dual to black holes and black
rings in the background \eqref{deconfmet:eq}. Unfortunately the
corresponding gravitational solutions have not yet been constructed;
however exact Black ring solutions to the vacuum Einstein equations
in 5 dimensions, were obtained in \cite{Emparan:2001wn} (see
\cite{Emparan:2006mm} for a review). These solutions were further
studied in \cite{Elvang:2006dd}. In this subsection we compare the
properties these black rings and black holes with our plasmaballs
and plasmarings, and find broad qualitative agreement between the
two.\footnote{While we expect the properties of plasmaballs and
plasmarings to match quantitatively with those of black holes and
black rings in the background \eqref{deconfmet:eq}, we could not
hope to find better than qualitative agreement with the properties
of the same objects in flat space.}

\begin{figure}
\begin{center}
  \small{A' - thin black ring, B' - thin black ring, thick black ring
  and black hole, C' - black hole.}\\
  \input{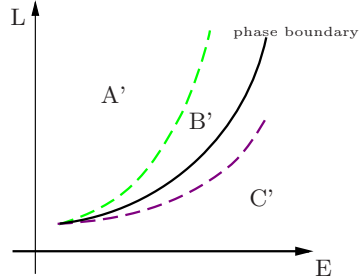}
  \caption{Existence regions and phase boundary for black holes /
rings.}\label{brphase:fig}
\end{center}
\end{figure}

In fig.\ref{brphase:fig} we have presented a schematic plot for the
existence of black hole and black ring solutions in 5 dimensional
flat space. This figure looks fairly similar to
figs.\ref{exist_final:fig},\ref{ent:fig}(b). The major qualitative
difference is the absence of the analogue of the region O (see fig.
\ref{exist_final:fig}) in fig. \ref{brphase:fig}. Thus unlike thin
black rings in flat 5 dimensional space, plasmarings (and so black
rings in Scherk-Schwarz AdS$_5$) have an upper bound to their
angular momentum at fixed energy.\footnote{This upper bound was
expected for black rings in AdS. The negative cosmological constant
has a similar effect to the dipole charge of \cite{Emparan:2004wy}.
We thank R.~Emparan for explaining this to us.}

It is interesting to pursue the comparison between these solutions
in more detail. The gravitational analogue of fig.\ref{ent:fig}
(presented as \cite[fig.3]{Emparan:2001wn}) looks fairly similar to
our figure. The main qualitative differences are: unlike for
plasmarings, the entropy of the large flat space black ring doesn't
go to zero at a finite angular momentum (it asymptotes to zero at
infinity) and the entropy of the small flat space black ring and
black hole go to zero at exactly the same point instead of the
slightly different values that we see. We expect that first of these
differences reflects a physical difference between black rings in
flat space and Scherk-Schwarz compactified AdS$_5$, the second
difference is an artefact of the breakdown of the fluid dynamics
approximation for extremal small rings (whose inner radius is always
of order the mean free path).

In even greater detail, we could quantitatively compare the
boundaries between regions O, A, B and C (see
fig.\ref{exist_prelim:fig}). These curves, as well as the phase
boundary, may be parameterised by $L=xE^y$ at large energies.

For black holes and black rings in flat space
$y_{AB}=y_{OC}=y_{BO}=y_{phase}=\frac{3}{2}$. For our plasmaballs
and plasmarings, as one can see in fig.\ref{logboundary:fig} (or
from (\ref{ballexist:eq}-\ref{thickexist:eq}) for the first three),
for large energy, we get $y_{OA}=2$, $y_{AB}=y_{OC}=\frac{3}{2}$,
$y_{BO}=1.25$ and $y_{phase}=1.25$ (see table \ref{bndscale:tab}).

It is meaningless to compare the $x$'s directly, as they are
dimensionful quantities. However, when two $y$'s have the same
value, the ratio of the corresponding $x$'s is dimensionless and may
be compared. For black rings $x_{AB}=\sqrt{32G/27\pi}$,
$x_{BO}=\sqrt{G/\pi}$ and $x_{phase}=\sqrt{256G/243\pi}$, so
$x_{AB}/x_{OC}=1$, $x_{OC}/x_{BO}=\sqrt{32/27}$ and
$x_{BO}/x_{phase}=9\sqrt{3}/16$. For plasmaballs and plasmarings, if
we used the dimensionless quantities \eqref{redch:eq}, we find
$x_{OA}=\frac{1}{16}$, $x_{AB}=x_{OC}=2/3^{3/2}$, $x_{BO}\approx
0.60$ and $x_{phase}\approx 0.67$. Therefore $x_{AB}/x_{OC}=1$ and
$x_{BO}/x_{phase}\approx 0.91$.

\begin{figure}
 \begin{center}
   \input{lglg.tpx}
   \input{grlg.tpx}
   \input{xf.tpx}
 \caption{(a) log-log plots of the boundaries, (b) gradients of
log-log plots, (c) $\tL/\tE^y \ra x$.}\label{logboundary:fig}
 \end{center}
\end{figure}


This is summarised in table \ref{bndscale:tab}. Note that the
extremality boundaries, OA, AB and OC, occur precisely where at
least one of the approximations discussed in \S\S\ref{sec:validity}
breaks down. Therefore, nothing quantitative about these boundaries
should be trusted.

\begin{table}
  \centering
  \begin{tabular}{|c|c|c|}
    \hline
    Quantity & Black rings & Plasmarings \\
    \hline
    $y_{OA}$ & N/A & 2 \\
    $y_{AB}$ & 3/2 & 3/2 \\
    $y_{OC}$ & 3/2 & 3/2 \\
    $y_{BO}$ & 3/2 & 1.25 \\
    $y_{phase}$ & 3/2 & 1.25 \\
    \hline
    $x_{AB}/x_{OC}$ & 1 & 1 \\
    $x_{OC}/x_{BO}$ & $\sqrt{32/27}$ & N/A \\
    $x_{BO}/x_{phase}$ & $9\sqrt{3}/16 \approx 0.97$ & 0.91 \\
    \hline
  \end{tabular}
  \caption{Comparison of scalings of boundaries for black rings and
plasmarings.}\label{bndscale:tab}
\end{table}


\subsection{Turning point stability}\label{sec:turn}

We have seen in \S\S\ref{sec:globstab} that the spinning plasma
solution of maximal entropy is the plasmaball (at low angular
momentum) or the large plasmaring (at high angular momentum). The
`phase transition' between these two solutions may be thought of as
being of first order (in the sense that the two competing solutions
are different at the phase transition point). The small plasmaring
is entropically subdominant to both the plasmaball and the large
plasmaring whenever it exists.

This situation appears to lend itself to a description in terms of a
Landau diagram, with the entropy given by a function of the
(unidentified) order parameter that has two maxima (the plasmaball
and the large plasmaring) separated by a single minimum (the small
plasmaring). This analogy suggests - and we conjecture that - the
small plasmaring is always dynamically unstable, while the
plasmaball and large plasmarings are dynamically stable with respect
to axisymmetric fluctuations.

An honest verification of our conjecture would require a study of
the spectrum of linear fluctuations about our plasmaball and
plasmaring solutions, an analysis that we have not carried out. In
this subsection, however, we present some evidence for our
conjecture, using the `turning point' stability analysis of
\cite{Poincare-lequdunemassflui:85} (see \cite{Arcioni:2004ww} for
discussion and references).

Consider a (not necessarily stable) equilibrium configuration that
changes from being stable to unstable under continuous variation.
The configurations we apply these considerations to are plasmarings;
according to our conjecture these rings are stable to axisymmetric
fluctuations when large but become unstable to the same modes when
small. At the boundary of stability, the matrix of second
derivatives of the entropy with respect to off shell variations (or
`order parameters') of the configuration under question develops a
zero eigenvalue. In the neighbourhood of this special point, a small
change in the thermodynamic potentials of the solution give rise to
a large change in the order parameter along the zero eigenvalue
direction (as such a change is entropically inexpensive). As argued
in \cite{1978MNRAS.183..765K, 1979MNRAS.189..817K, Sorkin:1981jc,
Sorkin:1982ut}, this results in a divergent contribution to the
second derivative of the equilibrium entropy as a function of
equilibrium thermodynamic quantities (for instance the angular
momentum at fixed energy) proportional to the negative inverse of
the small eigenvalue.

It follows that a configuration that changes stability has divergent
second derivatives of entropy with respect to - say - angular
momentum. Moreover the sign of this second derivative is positive in
the `more stable' phase and negative in the `less stable' phase.
Note that the turning point method gives information about the
change in the number of unstable directions about a solution, but
does not yield information about the absolute number of
instabilities.\footnote{Moreover, this method only links vertical
tangents - and not vertical asymptotes - in the graph of the first
derivative of entropy with respect to (say) angular momentum vs.\
angular momentum (a conjugacy diagram) to instabilities, as vertical
asymptotes occur at boundaries of equilibrium solution space instead
of separating solutions of differing degrees of stability.}

The turning point method is useful because it yields information
about stability properties, with respect to off shell fluctuations,
of phases, using information only about on shell variations. It is
especially useful in the study of nonextensive systems like black
holes, for which negative specific heats do not necessarily imply
dynamical instability (note that we're working with the microcanonical
ensemble, unlike the grand-canonical considerations of
\cite{Astefanesei:2005ad}). This method has been used to study the
stability of black rings in 5 dimensions \cite{Arcioni:2004ww,
Arcioni:2005fm}; it suggests that small black rings are always
unstable, while large black rings are more stable in that context.
This result corroborates the explicit linear fluctuation analysis
about the flat space black rings \cite{Elvang:2006dd}.

We now proceed to apply the turning point method to our plasmarings.
Define
\begin{align}
\label{betadef:eq}
  \beta &= \pdiffc[\tL]{\tS}{\tE} = \frac{1}{\tT} =
  \frac{1}{[g_+(\vo,\tw)]^{1/4}}\,,\\
\label{psidef:eq}
  \psi &= \pdiffc[\tE]{\tS}{\tL} = -\frac{\tO}{\tT} =
  -\frac{\tw}{[g_+(\vo,\tw)]^{1/4}}\,.
\end{align}

\begin{figure}
 \begin{center}
  \input{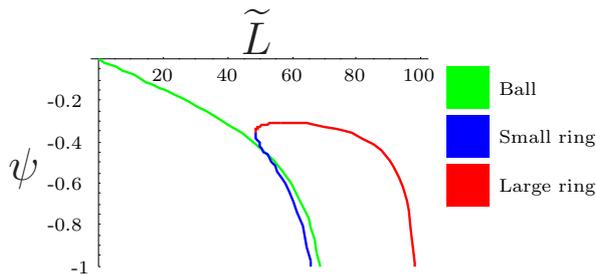}
 \caption{Conjugacy diagram: $\psi$ as a function of angular momentum, $\tL$
for fixed energy, $\tE=40$.}\label{turn:fig}
 \end{center}
\end{figure}

In fig.\ref{turn:fig} we have plotted $\psi$ against angular
momentum at fixed energy for our ring solutions. This graph has a
single turning point, precisely at the point at which the large ring
turns into a small ring. The slope of the curve turns from positive
(for the large ring) to negative (for the small ring) in upon
passing through the turning point, consistent with our conjecture
about the stability properties of plasmarings. More generally,
fig.\ref{turn:fig} is qualitatively similar to the equivalent graph
of \cite[fig.6(b)]{Arcioni:2004ww} for black hole and black rings in
flat 5 dimensional space, except that the large black ring curves
back down as we increase $\tL$. This difference has no impact on
stability analysis, as the turning point method links instabilities
to vertical tangents rather than horizontal tangents (even though a
heat capacity/susceptibility changes sign as one crosses a
horizontal tangent).

In conclusion, the turning point method indicates that the small
ring has an additional instability as compared to the large ring.
Note that it is perfectly possible that both the large and the small
ring are unstable, for example to fluctuations that break rotational
symmetry.


\section{Four dimensional plasmarings}\label{sec:4dim}

In the rest of this paper we turn to a consideration of localised
plasma configurations in certain massive 4 and 5 dimensional field
theories obtained by compactifying related 5 and 6 dimensional CFTs
on a Scherk-Schwarz circle. Although the field theories in question
are not gauge theories (e.g. the 5 dimensional massive theory could
be obtained by compactifying the (0,2) theory on the world volume of
an M5 brane on a Scherk-Schwarz circle) they undergo first order
`deconfining' transitions and the high temperature phase of these
theories admits a fluid dynamical description. The fluid
configurations we will construct are dual to localised black holes
and black rings in Scherk-Schwarz compactified AdS$_6$ and AdS$_7$
respectively.

In this section we study stationary solutions of fluid dynamics in 3+1
dimensional field theories. Our study will be less thorough than our 3 dimensional
analysis above; we find solutions analogous to those in 3 dimensions, but we postpone the
complete parameterisation and study of the thermodynamic properties
of these solutions to future work. In Appendix \ref{sec:5dim}
we have derived the equations relevant to stationary fluid flow in 5 dimensions, but
we leave the study of their solutions (and their higher dimensional counterparts)
to future work.

%

\subsection{Stress tensor and equations of motion}\label{sec:eom4d}

In this section we set up the equations of motion of our fluid. We
proceed in direct imitation of our analysis of $d=3$ above. We use
the metric
\begin{equation}\label{metric4d:eq}
  \dr s^2 = -\dr t^2 + \dr r^2 + r^2 \dr \phi^2 + \dr z^2\,.
\end{equation}
This gives the same non-zero Christoffel symbols as before
\eqref{chrst:eq}. We choose the origin so that $r=0$ is the axis of
rotation and there is a reflection symmetry in the plane $z=0$.


For our configurations, $u^\mu = \gamma(1,0,\omega,0)$ with
$\gamma=\prn{1-\omega^2r^2}^{-1/2}$. We assume that the surface can
be described by $f(r,z) = z- h(r)$. In the interior of the
fluid,This leads to the stress tensor
\begin{equation}\label{blkstr4d:eq}
 T^{\mu\nu}_\mathrm{perfect} =
 \begin{pmatrix}
  \gamma^2(\rho+\omega^2r^2P) & 0 & \gamma^2\omega(\rho+P)                  & 0 \\
  0                           & P & 0                                       & 0 \\
  \gamma^2\omega(\rho+P)      & 0 & \frac{\gamma^2}{r^2}(\omega^2r^2\rho+P) & 0 \\
  0                           & 0 & 0                                       & P \\
 \end{pmatrix}
\end{equation}
and the surface stress tensor
\begin{equation}\label{srfstr4d:eq}
 T^{\mu\nu}_\mathrm{surface} =
   \frac{\sigma \delta(z- h(r))}{\sqrt{1+h'(r)^2}}
 \begin{pmatrix}
  1+h'(r)^2 & 0      & 0                      & 0        \\
  0         & -1     & 0                      & -h'(r)   \\
  0         & 0      & -\frac{1+h'(r)^2}{r^2} & 0        \\
  0         & -h'(r) & 0                      & -h'(r)^2 \\
 \end{pmatrix}
\end{equation}

Just as in $d=3$, the only potentially nonzero term in
$T_\mathrm{dissipative}^{\mu \nu}$ is proportional to
$\diff{}{r}\brk{\frac{\tloc}{\gamma}}$. As in $d=3$, it will turn
out that this quantity vanishes on our solutions, so we simply
proceed setting $T^{\mu \nu}_{dissipative}$ to zero.


The equations of motion, $\nabla_\mu T^{\mu\nu}=0$, reduce to
\begin{equation}\label{eom4d:eq}
 \begin{split}
   0 &= \pdiff{P}{r} -\frac{\omega^2r}{1-\omega^2r^2}\,(\rho+P)
              \mp 2\sigma H h'(r)\, \delta(z-h(r))\,,\\
   0 &= \pdiff{P}{z} \pm 2\sigma H\, \delta(z-h(r))\,,
 \end{split}
\end{equation}
where the upper sign refers to the upper ($z>0$) surface and
\begin{equation}\label{meancurv:eq}
  H = \mp\frac{rh''+ h'(1+h'^2)}{2r(1+h'^2)^{3/2}}
\end{equation}
is the mean curvature of the surface \cite{Weisstein-SurfRevo:99}.

\subsection{Solutions}\label{sec:sol4d}

Our analysis so far has been rather general; to proceed further we
use the equations of state \eqref{thermint:eq}. We define
dimensionless variables as before
\begin{equation}\label{newvars4d:eq}
    \tw = \frac{\sigma\omega}{\rz} \,,   \qquad
    v = \omega r  \,,  \qquad
    \g(v) = \omega h(r)\,.
\end{equation}

Using the equation of state \eqref{thermint:eq}, we can rewrite
\eqref{eom4d:eq} in the bulkinterior of the fluid as
\begin{equation}\label{eombulk4d:eq}
 \begin{split}
  \frac{1}{\rho-\rz}\diff{\rho}{v} &=
                        \frac{5v}{1-v^2}\,,\\
                        \implies&
  \prn{\rho(v)-\rz}\prn{1-v^2}^{5/2} = 4K\rz\,,
 \end{split}
\end{equation}
where $K$ is an integration constant. This means that the pressure
and temperature are
\begin{equation}\label{PT4d:eq}
    P = \rz \prn{\frac{K}{(1-v^2)^{5/2}}-1}\,, \qquad
    T = \gamma \prn{\frac{K\rz}{\eos}}^{1/5}\,,
\end{equation}
(notice that this justifies our neglect of heat flow).

Integrating \eqref{eom4d:eq} across an outer surface gives
\begin{equation}\label{bcnd4d:eq}
  P = 2\sigma H \quad \text{or} \quad
  \frac{K}{(1-v^2)^{5/2}}-1  =
    -\tw \frac{v\g''+ \g'(1+\g'^2)}{v(1+\g'^2)^{3/2}}\,.
\end{equation}
%

This can be integrated once to give
\begin{equation}\label{bcyx4d:eq}
  \frac{v\g'}{\sqrt{1+\g'^2}} =
    -\frac{K}{3\tw(1-v^2)^{3/2}} + \frac{v^2}{2\tw}
     + \frac{C}{\tw}\,,
\end{equation}
where $C$ is another integration constant.

If we introduce a parameter $l$ equal to the distance along the
surface, measured from $(v,\g)=(\vo,0)$, we have
$\diff{l}{v}=-\sqrt{1+\g'^2}$. Then \eqref{bcyx4d:eq} can be written
as
\begin{equation}\label{intrinsic:eq}
\begin{split}
  \diff{\g}{l} &= \frac{2K-3(v^2+2C)(1-v^2)^{3/2}}
                            {6\tw v(1-v^2)^{3/2}}\,,\\
  \diff{v}{l} &= -\frac{\sqrt{36\tw^2v^2(1-v^2)^3
                           - \brk{2K - 3(v^2+2C)(1-v^2)^{3/2}}^2}}
                     {6\tw v (1-v^2)^{3/2}}\,,\\
  \diff{\g}{v} &= -\frac{2K-3(v^2+2C)(1-v^2)^{3/2}}
                     {\sqrt{36\tw^2v^2(1-v^2)^3
                           - \brk{2K - 3(v^2+2C)(1-v^2)^{3/2}}^2}}\,.\\
\end{split}
\end{equation}
It follows that the outer surface of our plasma configuration is
given by
\begin{equation}\label{profile:eq}
  \g(v)= \int_{\vo}^v \!\dr x \left(  -\frac{2K-3(x^2+2C)(1-x^2)^{3/2}}
                     {\sqrt{36\tw^2x^2(1-x^2)^3
                           - \brk{2K - 3(x^2+2C)(1-x^2)^{3/2}}^2}}
                           \right)
\end{equation}
Of course this only makes sense provided
\begin{equation}\label{sense:eq}
  6\tw x(1-x^2)^{3/2} \geq \abs{2K - 3(x^2+2C)(1-x^2)^{3/2}}
  \quad \forall\; x \in (v, v_0).
\end{equation}
Note also the conditions $\rho>\rz \implies K>0$ and, of course,
$0<\vi<\vo$.

Inner boundaries to the plasma configuration (if they exist) obey
the equation $P=-2\sigma H$. The equivalent of \eqref{intrinsic:eq},
with a new integration constant $D$ replacing $C$ (the integration
constant $K$ is a property of the plasma, not the surfaces), is
\begin{equation}\label{innerbnd:eq}
\begin{split}
  \diff{\g}{l} &= -\frac{2K-3(v^2+2D)(1-v^2)^{3/2}}
                            {6\tw v(1-v^2)^{3/2}}\,,\\
  \diff{v}{l} &= -\frac{\sqrt{36\tw^2v^2(1-v^2)^3
                           - \brk{2K - 3(v^2+2D)(1-v^2)^{3/2}}^2}}
                     {6\tw v (1-v^2)^{3/2}}\,,\\
  \diff{\g}{v} &= \frac{2K-3(v^2+2D)(1-v^2)^{3/2}}
                     {\sqrt{36\tw^2v^2(1-v^2)^3
                           - \brk{2K - 3(v^2+2D)(1-v^2)^{3/2}}^2}}\,.\\
\end{split}
\end{equation}
The profiles of such boundaries may be obtained by integrating the
equation above.


Even before doing any analysis, we will find it useful to give names
to several easily visualised, topologically distinct fluid
configurations.

%
%
\begin{description}
  \item[Ordinary ball:] $v'(l)=\g(l)=0$ at $v=\vo$. $\g'(l)>0$ for
  $0<v<\vo$. $\g'(l)=0$ at $v=0$.
  \item[Pinched ball:] $v'(l)=\g(l)=0$ at $v=\vo$. $\g'(l)>0$ for
  $0<v<\vm$. $\g'(l)=0$ at $v=\vm$. $\g'(l)<0$ for $0<v<\vm$.
  $\g'(l)=0$ at $v=0$.\footnote{Black holes with wavy horizons in six dimensions and above were predicted in \cite{Emparan:2003sy}.}
  \item[Ring:] $v'(l)=\g(l)=0$ at $v=\vo$. $\g'(l)>0$ for $\vm<v<\vo$.
  $\g'(l)=0$ at $v=\vm$. $\g'(l)<0$ for $\vi<v<\vm$. $v'(l)=\g(l)=0$
  at $v=\vi$, where $\vi<\vm<\vo$.
\end{description}
Examples of these surfaces can be seen in
figs.\ref{prof:fig}-\ref{surf:fig}. Each of these solutions could
have lumps of fluid eaten out of them. We will use the terms
\begin{description}
  \item[Hollow ball:] A ball (pinched or ordinary) with a ball cut out from its inside.
  \item[Hollow ring:] A ring with a ring cut out from its inside.
  \item[Toroidally hollowed ball:] A ball with a ring cut out from its inside.
\end{description}

It is easy to work out the horizon topology of the gravitational
solutions dual to the plasma topologies listed above.
\cite{Galloway:2005mf} have obtained a restriction on the topologies
of horizons of stationary black holes in any theory of gravity that
obeys the dominant energy condition; any product of spheres obeys
the conditions from their analysis.  Although the dominant energy
condition is violated in AdS space, in table \ref{topology:tab}, we
have listed all 4 dimensional horizons that are topologically
products of lower dimensional spheres, and note that all but one of
these configurations is obtained from the dual to plasma objects
named above ($B^3$ is a ball, $B^2$ is a disc and $B^1$ is an
interval). The last one, $T^4 = S^1\times S^1\times S^1\times S^1$,
is a marginal case of the theorem.

In the rest of this section we will determine all stationary,
rigidly spinning solutions of the equations of fluid dynamics
described above.

\begin{table}
  \begin{center}
  \begin{tabular}{|l|l|l|}
    \hline
    Horizon topology & Plasma topology & Object \\
    \hline
    $S^4$                     & $B^3$                     & Ball \\
    $S^3\times S^1$           & $B^2\times S^1$           & Ring \\
    $S^2\times S^2$           & $B^1\times S^2$           & Hollow ball \\
    $S^2\times S^1\times S^1$ & $B^1\times S^1\times S^1$ & Hollow ring \\
    $S^1\times S^1\times S^1\times S^1$
                              & None                      & None \\
    \hline
  \end{tabular}
  \end{center}
  \caption{Topologies of gravity and plasma solutions}\label{topology:tab}
\end{table}

\subsubsection{Ordinary ball}\label{sec:oball4d}

We search for solutions of \eqref{profile:eq} for which $\g'(v)$
vanishes at $v=0$ and blows up at the outermost point of the surface
$v_0$; we also require that $\g$ decrease monotonically from $0$ to
$\vo$. The first condition sets $K=3C$.  The condition that $v'(l)$
is zero at $\vo$ may be used to determine $\tw$ as a function of
$\vo$ and $K$ from the linear equation
\begin{equation}\label{oballcnstr:eq}
  2K - (3\vo^2+2K)(1-\vo^2)^{3/2} = 6\tw\vo(1-\vo^2)^{3/2} \,,
\end{equation}
(the choice of positive square root comes from the fact that the LHS
above is positive).

Note that the numerator of the formula for $\g'(v)$ be written as
\begin{equation*}
  2\brk{1-(1-v^2)^{3/2}}\prn{K-\frac{3v^2(1-v^2)^{3/2}}{2\brk{1-(1-v^2)^{3/2}}}}
\end{equation*}
and $2\prn{1-(1-v^2)^{3/2}}\geq3v^2(1-v^2)^{3/2}$. Thus, $K>1$
guarantees our monotonicity requirement. From \eqref{PT4d:eq}, we
see that this also ensures that the pressure is positive throughout
the ball.

In summary, the full set of ordinary ball solution is obtained by
substituting $C=K/3$ and $\omega= \omega(K, \vo)$ (obtained by
solving \eqref{oballcnstr:eq}) into \eqref{profile:eq}. This
procedure gives us a ball solution for any choice of $K>1$ and
$\vo>0$.

In figs.\ref{prof:fig},\ref{surf:fig} we present a plot of the profile $\g(v)$ for
the ball solution at $\vo=0.8$, $K=1.5$.

\subsubsection{Pinched ball}\label{sec:pball4d}

The pinched ball satisfies all the conditions of the ordinary ball
except for the monotonicity requirement on $\g(v)$; in fact the
function $\g(v)$ is required to first increase and then decrease as
$v$ runs from $0$ to $\vo$. It follows that $C$ and $\tw$ for these
solutions are determined as in the previous subsubsection ($C=K/3$
and $\omega $ from \eqref{oballcnstr:eq}) however the requirement
$\g''(v)>0$ at $v=0$ forces $K<1$. This ensures that $\g'(v)>0$ at
small $v$ and $\g'(v)<0$ at larger $v$. It also ensures that the
solution has negative pressure at the origin and positive pressure
at the outermost radius.

\begin{figure}[tbh]
\begin{center}
  \input{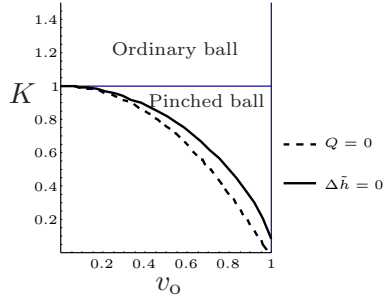}
  \caption{Domain of ball solutions.}\label{pbreg:fig}
\end{center}
\end{figure}

Not every choice of $(K,\vo)\in[0,1]$, however, yields an acceptable
pinched ball solution. As we decrease $\vo$ from 1, at fixed $K$, it
turns out that $\g(0)$ decreases, and in fact vanishes at a critical
value of $\vo$. Solutions at smaller $\vo$ are unphysical. The
physical domain,in $(K, \vo)$ space is given by the inequality
\begin{equation}\label{pballcnstr2:eq}
  \Delta \g \equiv -\!\!\int_0^{\vo} \diff{\g}{v}\,\dr v  =
  \int_0^{\vo}\!\!\!
      \frac{2K-(3v^2+2K)(1-v^2)^{3/2}}
           {\sqrt{36\tw^2v^2(1-v^2)^3-\brk{2K-(3v^2+2K)(1-v^2)^{3/2}}^2}}
      \,\dr v
  \geq 0\,.
\end{equation}
We should also ensure that \eqref{sense:eq} is not violated, i.e.
\begin{equation}\label{sensecnstr:eq}
  Q(\vo,K) \equiv \inf_{v \in (0,\vo)}\!
    \brc{36\tw^2 v^2(1-v^2)^{3} - \brk{2K - (3v^2+2K)(1-v^2)^{3/2}}^2}
    \geq 0\,.
\end{equation}
The boundary of the domain permitted by \eqref{pballcnstr2:eq}
is plotted in fig.\ref{pbreg:fig}. We have also plotted the boundary
of the region where \eqref{sensecnstr:eq} is violated. We see that
\eqref{pballcnstr2:eq} is the stricter constraint, and that
\eqref{sensecnstr:eq} is not violated for ordinary balls either. The
full set of pinched ball solutions is parameterised by values of
$\vo$ and $K$ in the region indicated in fig.\ref{pbreg:fig}.

In figs.\ref{prof:fig},\ref{surf:fig} we present an example of the profile $\g(v)$
for the pinched ball solution at parameters $\vo=0.8$, $K=0.55$.

\subsubsection{Ring}\label{sec:ring4d}

The plasma of the ring configuration excludes the region $v<\vi$; as
this region omits $v=0$,  $K$ and $C$ are not constrained as before.

\begin{figure}
\begin{center}
  \input{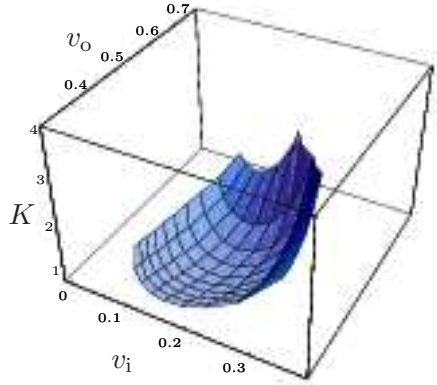}
  \caption{$K$ as a function of $\vi$ and $\vo$ for ring solutions.}\label{wplot:fig}
\end{center}
\end{figure}

As $v'(l)$ vanishes at $\vi,\vo$ we have the following constraints
\begin{equation}\label{ringcnstr:eq}
\begin{split}
  2K - 3(\vi^2+2C)(1-\vi^2)^{3/2} &= -6\tw\vi(1-\vi^2)^{3/2} \,, \\
  2K - 3(\vo^2+2C)(1-\vo^2)^{3/2} &= 6\tw\vo(1-\vo^2)^{3/2} \,.
\end{split}
\end{equation}
the choice of negative/positive square roots comes from the
requirements that $\g'(l)<0$ at $v=\vi$ and $\g'(l)>0$ at $v=\vo$.
These equations may be used to solve for $C$ and $\tw$ as a function
of $K, \vi, \vo$. $K(\vo,\vi)$ may then be determined from the
requirement that $\g(\vi)=\g(\vo)=0$, i.e.
\begin{equation}\label{ringcnstr2:eq}
  \int_{\vi}^{\vo} \diff{\g}{v} \, \dr v =
  -\int_{\vi}^{\vo}\!\!\!\frac{2K-3(v^2+2C)(1-v^2)^{3/2}}
    {\sqrt{36\tw^2v^2(1-v^2)^3-\brk{2K - 3(v^2+2C)(1-v^2)^{3/2}}^2}}
  \, \dr v
  = 0\,.
\end{equation}
In practice, it is easier to first eliminate $K$ and $C$ using
\eqref{ringcnstr:eq}, then substitute $\vi=\tw\zi$, $\vo=\tw\ro$ and
use \eqref{ringcnstr2:eq} to solve for $\tw$ at fixed $\zi$ and
$\zo$. after this, one can determine $K$, $\vi$ and $\vo$ from
$\tw$, $\zi$ and $\zo$. We present a 3 dimensional plot of $K$ as a
function of $\vi$ and $\vo$ for $1<\zo<10$, $0.1<\zi/\zo<0.9$ in
fig.\ref{wplot:fig}. All of these solutions have $K>0$, as required.
Unfortunately, with this method, one cannot see if there is a
physically acceptable solution for the whole range of $0<\vi<\vo<1$.
It appears that there is a solution for every value of $\zi<\zo$.

\begin{figure}
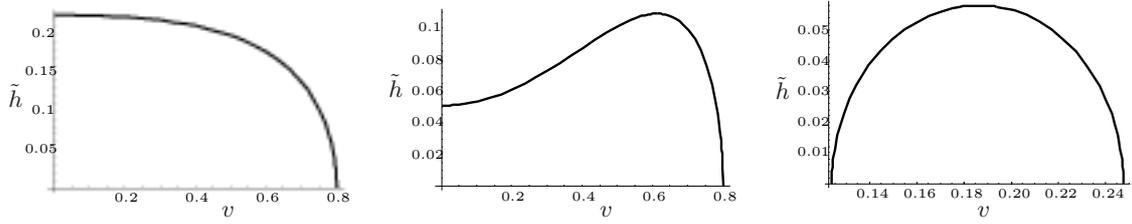

\begin{center}
  \input{obp.tpx}
  \input{pbp.tpx}
  \input{rp.tpx}
  \caption{Profile of the surface of an ordinary
ball, pinched ball and ring.}\label{prof:fig}
\end{center}
\end{figure}

\begin{figure}
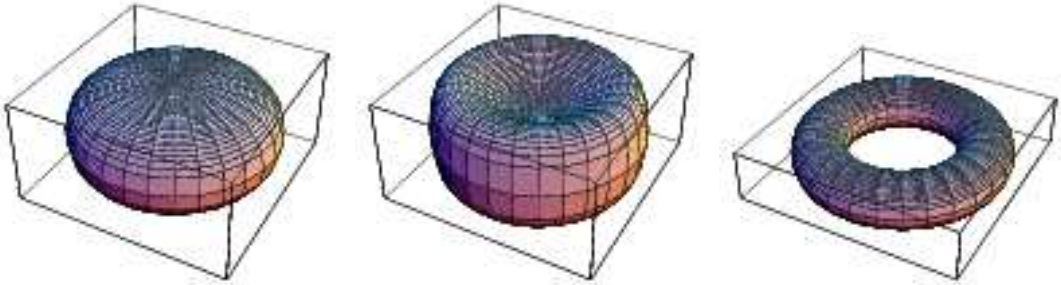

\begin{center}
  \input{obs.tpx}
  \input{pbs.tpx}
  \input{rs.tpx}
  \caption{3D plot of the surface of an ordinary
ball, pinched ball and ring.}\label{surf:fig}
\end{center}
\end{figure}

In figs.\ref{prof:fig},\ref{surf:fig} we plot the profile function $\g(v)$ for the
ring solution at parameters
$\zi=10$, $\zo=20$.


\subsubsection{Hollow ball}\label{sec:hball}

In this subsection we will demonstrate the non-existence of rigidly
rotating hollow ball solutions to the equations of fluid dynamics.
Let us suppose such a solution did exist. The inner surface must
have vanishing gradient at $v=0$; this sets $D=K/3$. Now let the
outermost point of the eaten out region be $v=\vot$. The inner
surface must have a vertical tangent at $\vot$. This also implies
that the outer surface also has a vertical tangent at $\vot$ (the
condition for a vertical tangent is identical for an outer or inner
surface). However, such points saturate the inequality
\eqref{sense:eq} and, as discussed in \S\S\S\ref{sec:pball4d}, this
never happens in the interior of a ball. It follows that hollow ball
solutions do not exist.

\subsubsection{Hollow ring and toroidally hollowed ball}\label{sec:hring}

Let us first consider the possibility of the existence of a toridally hollowed ball
solution. Let the innermost and outermost part of the hollowed out region occur at
$v=\vit$ and  $v=\vot$ respectively. Let us define $a(v)= 6 \tw v(1-v^2)^{{3/2}}$
and $b(v)=-2K+3(v^2+2D)(1-v^2)^{{3/2}}$ where $D$ is the integration constant for
the hollow. From \eqref{innerbnd:eq} it must be that
\begin{equation*}
  a(\vot)=b(\vot)\qquad
  a(\vit)=-b(\vit)\qquad
  |b(v)|<|a(v)|\; \forall   v \in (\vit, \vot)
\end{equation*}
For these conditions to apply, $b(v)$ must start out negative at $v=\vit$, increase, turn positive, and cut the $a(v)$ curve from below at $v=\vot$. We have performed a rough numerical scan of allowed values of parameters $(K, \tw, D)$; it appears that this behaviour never occurs
(although we do not, however, have a rigourous proof for this claim). For all physically
acceptable values of parameters, the curve $b(v)$ appears to either stay entirely below
$a(v)$ or to cut it from above.\footnote{We emphasise that this behaviour appears to be true only for $\tw>\tw_{min}(K)$ where $\tw_{min}(K)$ is the smallest allowed value of $\tw$ at fixed
$K$ (see fig.\ref{pbreg:fig}).  It is easy to arrange for $b(v)$ to cut $a(v)$ from below
when $\tw$ is taken to be arbitrarily small at fixed $K$ and $D$, but this is unphysical.}

These considerations, which could presumably be converted into a proof with enough effort,
lead us to believe that the existence of hollow balls is highly unlikely.
We believe that similar reasoning is likely to rule out the existence of hollow rings, although this is more difficult to explicitly verify, as our understanding of the parameter ranges for
acceptable ring solutions is incomplete.

In order to understand intuitively why hollow rings and toroidally hollow balls are unlikely,
note that the pressure at the inner and outermost parts of the hollowed out region is given by
\begin{equation*}
  P(\vit) = \rz\tw\prn{-|v''_{v=\vit}|+\frac{1}{\vit}}, \qquad
  P(\vot) = \rz\tw\prn{-|v''_{v=\vot}|-\frac{1}{\vot}},
\end{equation*}
where $v''_{v=\vit}$ is positive and $v''_{v=\vot}$ is negative.
Provided that $|v''_{v=\vit}|$ and $|v''_{v=\vot}|$ are  not
drastically different, we would require $P(\vit)>P(\vot)$. However,
the pressure increases monotonically with radius.


In conclusion, we strongly suspect, but have not yet fully proved,
that the full set of rigidly rotating solutions to the equations of fluid dynamics
in $d=4$ is exhausted by ordinary balls, pinched balls and rings.

%
%
%
%
%


%
%
%
%
%

\section{Discussion}\label{sec:discuss}

In this paper we have emphasised that the AdS/CFT correspondence
implies a duality between nonsingular classical gravitational
solutions with horizons, and solutions to the boundary equations of
fluid dynamics. This connection has previously been utilised by
several authors to obtain gravitational predictions for various
fluid viscosities and conductivities (see, for instance,
\cite{Son:2007vk} and references therein). The new element in our
work is the incorporation of boundaries separating the fluid from
the vacuum into the Navier-Stokes equations. This feature (which
relies on the explicit gravitational construction of the domain wall
in \cite{Aharony:2005bm}) allowed us to study stationary
\emph{finite energy} lumps of plasma, which are dual to localised
black holes and black rings in the bulk.

All our work (and easily imagined generalisations) apply to
confining field theories. Stationary black holes in such backgrounds
sit at the IR ends of the geometry; the boundary shadow of such
black holes is a lump of deconfined fluid of size $R + \CO
(\Lambda^{-1}_\mathrm{gap})$. The fluid dynamic equations accurately
describe such lumps only when $R  \gg \Lambda_\mathrm{gap}^{-1}$, in
the same limit the fluid yields an approximately local
representation of the horizon. Consequently, the AdS/CFT
correspondence provides an approximately local fluid description of
horizon dynamics in the long wavelength limit. This result is strongly
reminiscent of the Membrane paradigm of black hole physics
\cite{Thorne:1986iy, Parikh:1997ma,
Cardoso:2007ka}, and may constitute the precise version of this
idea in the context of asymptotically AdS spaces.

All the specific results of this paper are based on the equations of
state \eqref{thermint:eq}, which are valid for the high temperature
phase of Scherk-Schwarz compactified conformal field theories (dual
to gravity in Scherk-Schwarz compactified AdS space). However the
only qualitative feature of this equation of state that was
important for the existence of the solutions of this paper is that
the fluid pressure vanishes at finite energy density. In
fig.\ref{linvalidity:fig} we have plotted the 
maximum and minimum values of $\ln \tloc/\tc$ in our solutions,  as a function of
energy and angular momentum. Note that at large values of charges
(and away from extremality bounds) each of these quantities tends to
zero. This demonstrates that over large classes of our solutions,
the fluid temperature is always close to the phase transition
temperature. As a consequence such solutions `sample' only the fluid
equation of state only in the neighbourhood around the zero pressure
point, and so would exist in any fluid whose pressure vanishes at finite energy density.

Our results suggest several directions for future research. It would
be interesting to analyse the stability of small fluctuations about
the solutions presented in this paper. As we have mentioned in
\S\S\ref{sec:turn}, we expect the small ring to be unstable to
axisymmetric fluid fluctuations, while we expect the ball and the
large ring to be stable to such fluctuations. However, it is quite
possible that such an analysis would reveal that the large ring
solutions of this paper have a Plateau-Rayleigh type instability that
maps to Gregory-Laflamme instabilities (see also
\cite{Cardoso:2006ks})\footnote{We thank T. Wiseman for suggesting
this.} of the dual bulk solutions.

Although we have not mentioned this in the text, there exists a
scaling limit in which the thin plasmarings solutions simplify
greatly.\footnote{We thank T. Wiseman again for pointing this out to
us.} In this limit ($\tw \ra 0$ with $\vi$ fixed), the 3D plasmaring
reduces to a straight strip of moving fluid. The fluid
pressure vanishes on this strip, and the fluid velocity is constant
across the strip (more precisely $\vo = \vi +
\frac{1-\vi^2}{2\vi^2}\tw + \CO(\tw^2)$ so that $\zo-\zi =
\frac{1-\vi^2}{2\vi^2} + \CO(\tw)$). Similarly, there should exist
scaling limit under which the 4 dimensional plasmaring reduces to an
infinite stationary cylinder, with fluid flow along the axis.
Various dynamical properties of large rings (e.g. the potential
Gregory-Laflamme type instability alluded to in the previous
paragraph) will probably prove easiest to study in this scaling
limit.

It should also be relatively straightforward, and rather
interesting, to more fully analyse the thermodynamics of the four
dimensional solutions presented in this paper. This thermodynamics may
have interesting features; for  example, it has
been suggested that ultra-spinning black holes in six dimensional flat
space are unstable \cite{Emparan:2003sy} and it would be interesting
to see if the same is true of our (pinched?) plasmaballs.

An extension of our work to obtain the moduli space of five and higher
dimensional fluid configurations - and so seven and higher dimensional
gravitational black solutions should also be possible (though
analytic solutions may be harder to obtain in higher dimensions).
Such an extension would yield interesting information about horizon
topologies in higher dimensional gravitational theories. An obvious
conjecture based on intuition from fluid flows would be that the
full set of stationary fluid solutions in five dimensions appear in
three distinct topological classes; solutions whose bulk dual
topologies would be $S^5$, $S^4\times S^1$ and $S^3\times S^1\times
S^1$. The reason one might expect the last solution is that in five
(but no lower) dimensions, it is possible to have solutions that
rotate about two independent axes; these two rotations should be
able to create their own distinct centrifugal `holes', resulting in the
above topology. It would be
very interesting to check whether this conjecture is borne out.

%

\section*{Acknowledgements}

We would like to thank D.~Astefanesei, I.~Bena, K.~Damle,  R.~Emparan,
R.~Gopakumar, R.~Jena, S.~Raju and S.~Wadia for useful conversations.
We would especially like to thank O.~Aharony and T.~Wiseman for useful
comments. The work of S.M.\ was supported in part by a Swarnajayanti
Fellowship. We must also acknowledge our debt to the steady and generous
support of the people of India for research in basic sciences.

\section*{Appendices}
\appendix


\section{Five dimensional plasmarings}\label{sec:5dim}

In the bulk of this paper we  have presented an analysis of
stationary fluid configurations of the three and four dimensional fluid
flows. The analysis of analogous configurations in one higher
dimension has an interesting new element. The rotation group in four
spatial dimensions, $SO(4)$, has rank 2. Consequently a rotating
lump of fluid in five dimensions will be characterised by three rather than
two conserved charges (two angular momenta plus energy). When one of
the two angular momenta is set to zero, it seems likely that the set
of stationary solutions will be similar to those of the four
dimensional fluid; in this limit we expect ball and ring
configurations whose dual bulk horizon topologies are $S^5$ and
$S^4\times S^1$ respectively. However turning on the second angular
momentum on the ring solution could centrifugally repel the fluid
away from the second rotational axis, leading to a fluid
configuration with dual bulk horizon topology $S^1\times S^1\times
S^3$. Such configurations have not yet been discovered in gravity,
and it would be exciting to either construct them in fluid
mechanics, or to rule out their existence.

In this appendix we set up and partially solve the equations of
stationary fluid flow in five dimensions. While the stationary
equations of fluid dynamics are trivial to solve in the bulk in
every dimension, boundary conditions are harder to impose in higher
dimensions. In the particular case of 5 dimensions, the imposition
of these boundary conditions requires the solution of a 2nd order
ordinary differential equation, that we have not (yet?) been able to
solve. It may be that a full study of this case would require
careful numerical analysis, which we leave to future work. In the
rest of this appendix we simply set up the relevant equations, and
comment on the dual bulk interpretations of various possible
solutions.

Consider a fluid propagating in flat five dimensional space
\begin{equation*}
  \dr s^2 = -\dr t^2 + \dr r_1^2 + r_1^2 \dr \phi_1^2
              + \dr r_2^2 + r_2^2 \dr \phi_2^2\,.
\end{equation*}
Consider a fluid flow with velocities given by
$u^\mu=\gamma(1,0,\omega_1,0,\omega_2)$, where
$\gamma=(1-v_1^2-v_2^2)^{-1/2}$, $v_1=\omega_1r_1$ and
$v_2=\omega_2r_2$. Let the fluid surface be given by
$f(r_1,r_2)=r_2-h(r_1)=0$.

The stress tensor evaluated on such a fluid configuration is given
by
\begin{equation*}
\begin{split}
  T^{\mu\nu}_\mathrm{perfect} &=
    \begin{pmatrix}
      \gamma^2(\rho+(v_1^2+v_2^2)P) & 0 & \gamma^2\omega_1(\rho+P) & 0 & \gamma^2\omega_2(\rho+P) \\
      0 & P & 0 & 0 & 0 \\
      \gamma^2\omega_1(\rho+P) & 0 & \frac{\gamma^2}{r_1^2}(v_1^2\rho+(1-v_2^2)P) & 0 & \gamma^2\omega_1\omega_2(\rho+P) \\
      0 & 0 & 0 & P & 0 \\
      \gamma^2\omega_2(\rho+P) & 0 & \gamma^2\omega_1\omega_2(\rho+P) & 0 & \frac{\gamma^2}{r_2^2}(v_2^2\rho+(1-v_1^2)P) \\
    \end{pmatrix}
     \\
\end{split}
\end{equation*}
\begin{equation*}
\begin{split}
  T^{\mu\nu}_\mathrm{dissipative} &=-\kappa\gamma^2
    \begin{pmatrix}
      0        & \p_{r_1}         & 0                & \p_{r_2}         & 0 \\
      \p_{r_1} & 0                & \omega_1\p_{r_1} & 0                & \omega_2\p_{r_1} \\
      0        & \omega_1\p_{r_1} & 0                & \omega_1\p_{r_2} & 0 \\
      \p_{r_2} & 0                & \omega_1\p_{r_2} & 0                & \omega_2\p_{r_2} \\
      0        & \omega_2\p_{r_1} & 0                & \omega_2\p_{r_2} & 0 \\
    \end{pmatrix}\frac{\tloc}{\gamma}
     \\
  T^{\mu\nu}_\mathrm{surface} &=\frac{\sigma\delta(r_2-h(r_1))}{\sqrt{1+h'^2}}
    \begin{pmatrix}
      1+h'^2 & 0   & 0                     & 0     & 0 \\
      0      & -1  & 0                     & -h'   & 0 \\
      0      & 0   & -\frac{1+h'^2}{r_1^2} & 0     & 0 \\
      0      & -h' & 0                     & -h'^2 & 0 \\
      0      & 0   & 0                     & 0     & -\frac{1+h'^2}{r_2^2} \\
    \end{pmatrix}
\end{split}
\end{equation*}

As usual, we will temporarily ignore
$T^{\mu\nu}_\mathrm{dissipative}$, justifying this when we find that
$\tloc\propto\gamma$. The nontrivial equations of motion that follow
from \eqref{Epconsv:eq} take the form
\begin{equation}\label{5deom:eq}
\begin{split}
  0 = \nabla_\mu T^{\mu r_1}
   &= \p_{r_1}P -\gamma^2\omega_1^2r_1(\rho+P) -\sigma G h'(r_1)\delta(r_2-h(r_1))\\
  0 = \nabla_\mu T^{\mu r_2}
   &= \p_{r_2}P -\gamma^2\omega_2^2r_2(\rho+P) +\sigma G \delta(r_2-h(r_1))
\end{split}
\end{equation}
where
\begin{equation*}
  G = -\frac{r_1hh''+(1+h'^2)(hh'-r_1)}{r_1h(1+h'^2)^{3/2}}
\end{equation*}

Using the equation of state \eqref{thermint:eq}, the equations of
motion in the fluid interior
\begin{equation*}
  \pdiff{\rho}{v_1} = 3(\rho-\rz) \frac{2v_1}{1-v_1^2-v_2^2}  \qquad
  \pdiff{\rho}{v_2} = 3(\rho-\rz) \frac{2v_2}{1-v_1^2-v_2^2}
\end{equation*}
are easily solved and we find
\begin{equation*}
  (\rho-\rz)(1-v_1^2-v_2^2)^3 = 5K\rz \qquad
  P = \rz\prn{\frac{K}{(1-v_1^2-v_2^2)^3}-1}
  \qquad \tloc = \gamma\brk{\frac{K\rz}{\eos}}^{1/6}.
\end{equation*}
Integrating the equations of motion \eqref{5deom:eq} across the
surface we obtain the condition (the upper sign should be used for
upper surfaces)
\begin{equation}\label{5dsurf:eq}
  \frac{K}{(1-\omega_1^2r_1^2-\omega_2^2h^2)^3}-1 =
     \mp\frac{\sigma}{\rz}
     \frac{r_1hh''+(1+h'^2)(hh'-r_1)}{r_1h(1+h'^2)^{3/2}}
\end{equation}
%

Unfortunately we have not yet been able to solve this equation; we
postpone further analysis of \eqref{5dsurf:eq} to future work. In
the rest of this appendix we qualitatively describe possible types
of solutions to these equations, and their bulk dual horizon
topologies.

\begin{figure}
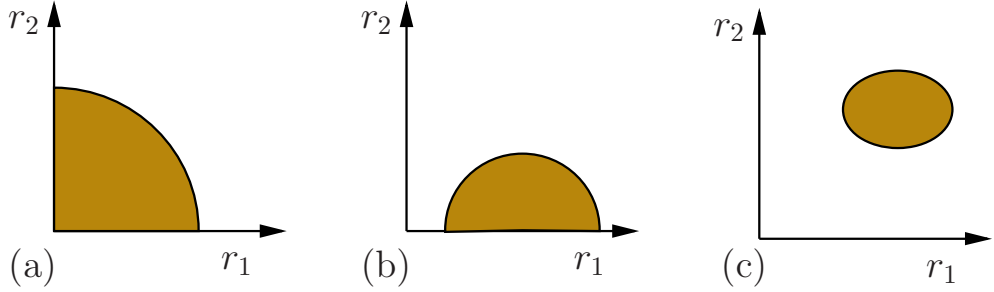

\begin{center}
  \input{5ball.tpx}\hspace{0.5cm}
  \input{5ring.tpx}\hspace{0.5cm}
  \input{5torus.tpx}
  \caption{Topologies of five dimensional solutions.}\label{5d:fig}
\end{center}
\end{figure}

In fig.\ref{5d:fig}, we have sketched some possible topologies for
these solutions. The first touches both the $r_1=0$ and $r_2=0$ axes
and we refer to this as a ball. The second type only touches one of
these axes and we refer to this as a ring. The third type touches
neither of the axes, as the plasma has the topology of a solid
three-torus we refer to this as a torus.

Each of these could be pinched near either axis, and there could be
hollow versions (though the considerations of \S\ref{sec:4dim} make
it seem unlikely that hollow configurations will actually be
solutions).

The horizon topology of the dual black object can be found by
fibering three circles over the shapes in fig.\ref{5d:fig}. One of
these circles degenerates at each axis (the angular coordinates
$\phi_1$ and $\phi_2$), and the other degenerates on the fluid
surface (the Scherk-Schwarz circle). The topologies generated are:
%
%
%
  \begin{center}
  \begin{tabular}{|l|l|l|}
    \hline
    Horizon topology & Plasma topology & Object \\
    \hline
    $S^5$                     & $B^4$                     & Ball \\
    $S^4\times S^1$           & $B^3\times S^1$           & Ring \\
    $S^3\times S^1\times S^1$ & $B^2\times S^1\times S^1$ & Torus \\
    $S^3\times S^2$           & $B^1\times S^3$           & Hollow ball \\
    $S^2\times S^2\times S^1$ & $B^1\times S^2\times S^1$ & Hollow ring \\
    $S^2\times S^1\times S^1\times S^1$
                              & $B^1\times S^1\times S^1\times S^1$
                                                          & Hollow Torus \\
    $S^1\times S^1\times S^1\times S^1\times S^1$
                              & None                      & None \\
    \hline
  \end{tabular}
  \end{center}


\section{Notation}\label{app:notation}

We work in the $(-++)$ signature. $\mu,\nu$ denote space-time
indices.

\vp
  \centering
  \begin{tabular}{||r|l||r|l||}
    \hline
    Symbol & Definition & Symbol & Definition \\
    \hline
    $\floc$ & Plasma free energy & $f$ & Free energy density \\
    $\eloc$ & Plasma energy & $\rho$ & Proper density \\
    $\sloc$ & Plasma entropy & $s$ & Proper entropy density \\
    $\tloc$ & Plasma temperature & $P$ & Pressure \\
    $\rz$ & Plasma vacuum energy & $\tc$ & Deconfinement temperature \\
    $\eos$ & see \eqref{freecnfng:eq} & $\rho_c$ & Deconfinement density \\
    \hline
%
    $T^{\mu\nu}$ & Stress tensor & $u^\mu$ & $\diff{x^\mu}{\tau}=\gamma(1,\vec{v})$ \\
    $\sigma$ & Surface tension & $\gamma$ & $\prn{1-v^2}^{-1/2}$ \\
    $f(x)$ & Surface at $f(x)=0$ & $\omega$ & Angular velocity \\
    $f_\mu$ & $\p_\mu f / \sqrt{\p f \cdt \p f}$ & $v$ & $\omega r$ \\
    $\theta,\sigma^{\mu\nu},a^\mu,P^{\mu\nu}$ & see \eqref{fluidtensors:eq}
       & $\zeta,\eta$ & Bulk, shear viscosity \\
    $q^\mu$ & Heat flux & $\kappa$ & Thermal conductivity \\
    \hline
  \end{tabular}

  \begin{tabular}{||r|l||r|l||}
    \hline
    Symbol & Definition & Symbol & Definition \\
    \hline
    $\tw$ & $\sigma\omega / \rz$ & $\z$ & $\rz r / \sigma$  \\
    $\ro$ & Outer radius & $\vo$ & $\omega \ro$ \\
    $\ri$ & Inner radius & $\vi$ & $\omega \ri$ \\
    $\zo$ & $\vo/\tw$ & $g_\pm(v)$ & Boundary conditions \eqref{ringsol:eq} \\
    $\zi$ & $\vi/\tw$ &  &    \\
    \hline
%
   $E$ & Ball/ring energy & $S$ & Ball/ring entropy \\
    $L$ & Angular momentum & $T$ & Ball/ring temperature \\
    $\Omega$ & Ball/ring angular velocity
       & $\Delta u$ & Validity criterion \eqref{validity2:eq}\\
    $\tE,\tL,\tS,\tT,\tO$ & See \eqref{redch:eq}
       & $\beta$ & $(\p\tS/\p\tE)_{\tL}$ \\
    O,A,B,C & Existence regions, see fig.\ref{exist_prelim:fig}
       & $\psi$ & $(\p\tS/\p\tL)_{\tE}$ \\
    $\tOr,\tA,\tB,\tC$ & Existence regions, see fig.\ref{exist_final:fig}
       & $x,y$ & Boundary scalings in \S\S\ref{sec:existcomp}  \\
    \hline
%
    $h$ & Surface height $z=h(r)$ & $\g$ & $\tw h$ \\
    $H$ & Mean curvature
      & $K,C,D$ & Integration constants in \S\ref{sec:4dim} \\
    $\Delta\g,Q$ & Consistency conditions
               (\ref{pballcnstr2:eq},\ref{sensecnstr:eq})
      & $G$ & Curvature in \S\ref{sec:5dim} \\
    \hline
  \end{tabular}

\bibliographystyle{utcaps}
\bibliography{plasmaring-minimal}

\end{document}